\documentclass[11pt]{article} 

\def\ket#1{ $ \left\vert  #1   \right\rangle $ } 
\def\ketm#1{  \left\vert  #1   \right\rangle   } 
\def\spr#1#2{ $ \left\langle \left. #1 \right\vert #2  \right\rangle $ } 
\def\sprm#1#2{  \left\langle #1 \left\vert \right. #2 \right\rangle   }

\def\proc#1{  {\bf #1}   \medskip} 
\def\subproc#1{  { #1}   \medskip} 
\def\procref#1{{\sf #1}} 
\def\procargop{ {\small\bf Argument options:} } 
\def\procadd{ {\small\bf Additional information:} } 
\def\procout{ {\small\bf Output:} } 
\def\procsee{ {\small\bf See also:} } 
 
\def\map{\hspace{-0.2mm}{\small\bf $\clubsuit$} }

\def\sixjm#1#2#3#4#5#6{  \left\{ \matrix{ #1 & #2 & #3  \cr 
                                          #4 & #5 & #6  } \right\}   }  

\def\ninejm#1#2#3#4#5#6#7#8#9{  \left\{ \matrix{ #1 & #2 & #3  \cr 
                                                 #4 & #5 & #6  \cr 
                                                 #7 & #8 & #9 } \right\}   } 

\def\jfactorm#1{  \left[ #1 \right] } 

\def\tmm#1{      \buildrel - \over #1  }  
  
\def\tpm#1{      \buildrel + \over #1  }  
\def\tpmm#1{     \buildrel \pm \over #1  }  
\def\plus#1{ \buildrel + \over #1} 
\def\minus#1{\buildrel - \over #1}

\begin{document} 
\sloppy 
 
\hyphenation{Ra-cah-expr}  
\hyphenation{Ra-cah-expr-}  
\hyphenation{Ra-cah-expres-sion}

\title{Maple procedures for the coupling of angular momenta.           \\ 
       VI. $LS-jj$ transformations} 
 
\author{G.\ Gaigalas${}^{\,a}$ and  S.\ Fritzsche${}^{\,b \, \dagger }$  \\ 
        \\ 
        $^a$ Institute of Theoretical Physics and Astronomy,           \\ 
        A.\ Go\v{s}tauto 12, Vilnius 2600, Lithuania.                  \\ 
        $^b$ Fachbereich Physik, Universit\"a{}t Kassel,               \\ 
        Heinrich--Plett--Str. 40, D--34132 Kassel, Germany.            \\   
        ${}^\dagger$ e--mail: s.fritzsche@physik.uni-kassel.de         \\ 
        \\ 
        \\ 
        }

\maketitle 
 
\date{} 
 
\begin{abstract} 
Transformation matrices between different coupling schemes are required, 
if a reliable classification of the level structure is to be obtained for 
open--shell atoms and ions. While, for instance, relativistic computations 
are traditionally carried out in $jj-$coupling, a $LSJ$ coupling notation 
often occurs much more appropriate for classifying the valence--shell 
structure of atoms. Apart from the (known) transformation of single open 
shells, however, further demand on proper transformation coefficients has 
recently arose from the study of open $d-$ and $f-$shell elements, the 
analysis of multiple--excited levels, or the investigation on inner--shell 
phenomena. Therefore, in order to facilitate a simple access to 
$LS \,\leftrightarrow\, jj$ transformation matrices, here we present an 
extension to the \textsc{Racah} program for the set--up and the transformation 
of symmetry--adapted functions. A flexible notation is introduced 
for defining and for manipulating open--shell configurations
at different level of complexity which can be extended also to other coupling 
schemes and, hence, may help determine an \textit{optimum} classification 
of atomic levels and processes in the future.
\end{abstract} 
 
\bigskip 
\bigskip 
 
\hspace{0.84cm} PACS: 3.65F, 2.90+p.

\newpage 
 
{\large\bf PROGRAM SUMMARY} 
 
\bigskip

{\it Title of program:} \textsc{Racah} 
 
\bigskip

{\it Catalogue identifier:}  ADQP
 
\bigskip 
 
\textit{Program Summary URL:} http://cpc.cs.qub.ac.uk/summaries/ADQP
 
\bigskip  
 
\textit{Program obtainable from:} CPC Program Library,  
     Queen's University of Belfast, N.\ Ireland.  
 
\bigskip

\textit{Licensing provisions:} None. 
 
\bigskip

\textit{Computers for which the program is designed:}   \newline 
     All computers with a license of the computer algebra  
     package \textsc{Maple} [1]. 
 
\bigskip

\textit{Installations:} University of Kassel (Germany). 
 
\bigskip

\textit{Operating systems under which the program has been tested:}   
     Linux 6.1+. 
 
\bigskip

\textit{Program language used:} \textsc{Maple} V, Release 6 and 7. 
 
\bigskip

\textit{Memory required to execute with typical data:}  30 MB. 
 
\bigskip

 
\bigskip

\textit{Nature of the physical problem:}  \newline 
     For open--shell atoms and ions, a reliable classification of the level 
     structure often requires the knowledge of the $LS-jj$ transformation 
     matrices in order to find the main components of the wave functions 
     as well as their proper spectroscopic notation. Apart from the 
     transformation of individual (sub--) shell states, matrices of much 
     larger complexity arise for the transformation of symmetry--adapted 
     configuration state functions which are constructed from the coupling
     of two or more open shells. 
      
\bigskip

\textit{Method of solution:}  \newline 
     $LS-jj$ transformation matrices are provided for all (sub--) shell states 
     with orbital angular momenta $l \,\le\, 3$ in the framework of the  
     \textsc{Racah} program [2]. These matrices are then utilized to  
     transform symmetry--adapted configuration state functions (CSF), including 
     the coupling of two open shells. Moreover, a simple notation is  
     introduced to handle such symmetry functions interactively and to  
     transform even atomic states which are given as a superposition of CSF.

\bigskip

\textit{Restrictions onto the complexity of the problem:}  \newline  
     The program presently supports all shell states with $ l \,\le\, 3 $, 
     i.e.\ up to open $f-$shells, in $LS-$coupling and with 
     $ j \,\le\, 7/2 $, i.e.\ up to open $f_{7/2}-$ and $g_{7/2}-$subshells,  
     in $jj-$coupling. For the transformation of configuration state functions, 
     the coupling of two open $LS-$shells or, correspondingly, four  
     $jj-$subshells are also supported. In $jj-$coupling, however, a  
     \textit{standard order} [cf.\ subsection 2.2] is always assumed for the  
     coupling sequence of the individual shells. Several simplifications are 
     used on the basis of this standard order. 
      
\bigskip

\textit{Unusual features of the program:}  \newline  
     Apart from the ''interactive access'' to the $LS-jj$ transformation 
     matrix elements between (sub--) shell states in $LS-$ and $jj-$coupling,  
     a complete transformation of the coupling scheme can be carried out 
     also for configuration respectively atomic state functions, just by 
     typing a few lines at \textsc{Maple}'s prompt. 
     To simplify the handling of the program, a short but very powerful 
     notation has been introduced which help the user to \textit{construct} 
     stepwise symmetry--adapted functions of different complexity. 
     But although the program presently supports only shell states in $LS-$ 
     and $jj-$coupling, the same notation can be extended also to incorporate 
     further coupling schemes in the future. 
     The main commands of the present extension are described in detail in 
     Appendix B; for a quick reference on the current capabilities of the 
     \textsc{Racah} program, we refer the reader to Ref.\ [3] and to a list  
     of all available commands in the file \texttt{Racah-commands.ps} which 
     is appended to the code. 
      
\bigskip

\textit{Typical running time:} 
     The program replies promptly on most requests. Even large 
     tabulations of $LS-jj$ transformation matrices can be carried out  
     in a few (tens of) seconds.

\smallskip 
 
\begin{flushleft} 
{\it References:}   \\{} 
     [1] Maple is a registered trademark of Waterloo Maple Inc.        \\{} 
     [2] S. Fritzsche, Comp.\ Phys.\ Commun.\ {\bf 103}, 51 (1997);     
         G.\ Gaigalas, S.\ Fritzsche,  B.\ Fricke,  
         Comp.\ Phys.\ Commun.\ \textbf{135}, 219 (2001).              \\{} 
     [3] S. Fritzsche, T. Inghoff, T. Bastug and M. Tomaselli, 
         Comp.\ Phys.\ Commun.\ \textbf{139}, 314 (2001).           
\end{flushleft}

\newpage 
 
{\large\bf LONG WRITE--UP} 

\section{Introduction} 
 
The classification of the level structure of open--shell atoms and ions 
is a \textit{non--trivial} task which occurs frequently in the interpretation 
of complex spectra. In the analysis of optical spectra, for instance, the  
correct knowledge of the $LSJ$ spectroscopic notation of the atomic states 
may help isolate individual levels and terms without that the theoretical 
energies from ab--initio computations need be accurate
enough for a direct assignment of the observed lines. In fact, such a demand 
arises already for rather simple shell structures such as the 
spectrum of Ne~II \cite{Ne_II}, for which the lowest excited 
$2s 2p^{\,6} \;\: ^2S$ term occurs high--up in the theoretical level structure, 
even if a sizeable wave function expansion is applied, and therefore may lead 
to misassignments --- if no additional information about further properties 
of these levels or about their representation in different coupling schemes 
is available. Since, today, most relativistic computations are carried out 
in $jj-$coupling, an efficient and reliable $LS-jj$ transformation of atomic 
states is of primary interest. 
 
\medskip 

For atoms with a single open shell and, in particular with an open $s-$ or
$p-$shell, the $LS-jj$ transformation matrices are well known and can be
obtained from different sources \cite{Cowan:81,Rudzikas:97,Dyall:84}. 
These matrices are also the \textit{building blocks} for the transformation 
of all symmetry--adapted functions and are often simply abbreviated by
\begin{eqnarray} 
\label{LS-jj-subshell-me} 
   \sprm{l^{\,N} \; \alpha LS\ J}{ 
         \left( \tmm{\kappa}^{\tmm{N}} \tmm{\nu} \tmm{J}, \ 
                \tpm{\kappa}^{\tpm{N}} \tpm{\nu} \tpm{J} \right) J} \, ,
\end{eqnarray} 
i.e.\ in terms of Fourier coefficients of the corresponding shell states 
in the (re--coupled) basis. However, such matrices for a single--shell
configuration with fixed occupation $N \,=\, \tmm{N}\,+\,\tpm{N}$ are 
only of little help in transforming atomic or configuration states 
with a more complex shell structures for which the individual (sub--) shell
states need to be treated consistently with respect to their definition,
choice of quantum numbers as well as their phase relation to each other. 
Therefore, in order to extent the single--shell matrices 
(\ref{LS-jj-subshell-me}) to open $d-$ and $f-$shell configurations or to 
evaluate these transformation matrices for complex shell structures, 
insight into the construction of the subshell states is required for 
the all the coupling schemes under consideration. 
Moreover, the size of the transformation matrices (\ref{LS-jj-subshell-me})
increases rapidly with the orbital angular momentum $l$ due to the large
number of (allowed) projections $m_l$ of the electrons in any open--shell 
configuration $(n_1 l_1)^{N_1},\ (n_2 l_2)^{N_2},\ \dots $ 
\cite{Gaigalas/ZR:02}. 
For these two reasons and due to the complexity of the \textit{recoupling 
coefficients}, which arise in the evaluation of the transformation coefficients
[cf.\ Eqs. (\ref{general-rec-1}--\ref{general-rec-2})], these matrix elements 
are often not available from the literature, not to mention the efficiency 
of their use if more than one open shell is involved and if such 
transformation need to be carried out explicitly.
 
\medskip 

Today, an alternative and much simpler access to the transformation between 
different coupling schemes is possible by means of computer--algebraic 
manipulations. For the coupling of angular momenta, for instance, 
such a framework for symbolic manipulations have been developed by us during 
recent years and is now known as the \textsc{Racah} program 
\cite{Fritzsche:97}. This program is a powerful tool in simplifying formal 
expressions from the theory of angular momentum. Recent developments to this 
package concerned not only the fast and reliable evaluation of Racah expression
but also the implementation of standard quantities
\cite{Gaigalas:01}, spherical harmonics \cite{Inghoff/Fri:01} as well as 
the evaluation of recoupling coefficients \cite{Fri/Inghoff:01}. Therefore, 
the \textsc{Racah} package also meets (most of) the
\textit{basic requirements} which are needed for the transformation of general,
symmetry--adapted functions between different coupling schemes. With the 
present extension to the \textsc{Racah} package, we now support a convenient
set--up and application of the $LS-jj$ transformation matrices for all atomic
(sub--)shells with orbital angular momenta $ l \,\le\, 3$. In addition, 
a powerful notation is provided for dealing with symmetry--adapted functions
at different level of complexity such as atomic and configuration state
functions as obtained from relativistic computations. In the present
implementation, we support the transformation of such symmetry functions 
with up to two open (nonrelativistic) $LS-$shells or up to four (relativistic) 
$jj-$subshells, respectively. For even more complex shell structures, moreover,
we intent to utilize and implement these developments directly into
the available atomic codes such as \textsc{Grasp92} \cite{Grasp92} or the 
\textsc{Ratip} package \cite{Ratip}. But already with the present extension 
of the \textsc{Racah} program, a major step in the $LSJ$ classification of 
atomic and ionic levels has been achieved.
 
\medskip 
 
In the next section, we first explain the construction of symmetry--adapted 
functions in $LS-$ and $jj-$coupling, respectively, as well as the evaluation 
of the transformation matrices. This is followed in section 3 by a short review 
about \textsc{Racah}'s program structure and how it is distributed before we 
illustrate and discuss several examples in section 4. Apart from the
transformation of configuration states with a single open shell, our third 
example displays the transformation of two atomic levels as they may arise
in standard computations. Section 5 outlines the algebraic evaluation 
and simplification of the transformation matrices, of course, by making 
use again of \textsc{Racah} itself. This section points to the 
\textit{road} which we need to go in order to deal with general open--shell 
states and their transformation among different coupling schemes in the 
future. Finally, a few comments on further and highly desirable extensions 
of the present work are given in section 6.

\section{$LS-jj$ transformation matrices}

\subsection{Transformation of subshell states} 
 
For a successful transformation of symmetry--adapted functions from one 
coupling scheme to another, it is first necessary to understand the 
construction of these functions in some detail. In atomic shell theory, 
symmetry--adapted configuration states are usually constructed from 
antisymmetrized states of $N$ \textit{equivalent} electrons of a given
shell $(nl)$, to which we briefly refer as (sub--) shell states below.  
In $LS-$coupling, for example, such a subshell states of the shell $(nl)$
is written as~\cite{Rudzikas:97} 
\begin{equation} 
\label{eq:quasispin-LS-a} 
   \ketm{nl^{\,N} \; \alpha LS} 
\end{equation} 
where $\alpha $ represents all additional quantum numbers which,
apart from the total orbital angular momentum $L$ and total spin $S$, 
are needed for the unique classification of these states. In practise, 
an additional number $\alpha $ is needed only for subshells with orbital 
angular momenta $l \ge 3$, i.e.\ for electrons from the $f$--, $g$--, \ldots 
shells. A list of all possible subshell states for open $s$--, $p$--,  $d$-- 
and $f$--shells, both in quasispin and seniority notation, has been displayed 
previously in Ref.\ \cite{Gaigalas:01}, table 1.  For the subshell states 
(\ref{eq:quasispin-LS-a}), moreover, the angular momenta $L$ and $S$ can be 
coupled also to an total angular momentum $J$, \ket{nl^{\,N} \; \alpha LSJ}, 
which gives rise to the so--called $LSJ$ notation. Of course, further
{\em additional} intermediate angular momenta will arise if the subshell states
of two or more open shells are coupled to each other which, however, should not
be confused with the current discussion about the subshell states for 
{\em equivalent} electrons.
 
\medskip 
 
In $jj-$coupling, similarly, the subshell states of $N$ equivalent electrons 
of a subshell $(n\kappa)$ are represented by 
\begin{equation} 
\label{eq:quasispin-jj-a} 
   \ketm{n\kappa^{N} \; \nu J} \;  
\end{equation} 
where $\kappa$ is the relativistic (angular momentum) quantum number  
\begin{eqnarray} 
   \kappa & = & \pm \, (j+1/2) \qquad \mbox{for} \qquad l \;=\; j \pm 1/2 \;  
\end{eqnarray} 
and two further quantum numbers $\nu$ and $J$ are found sufficient to classify 
all subshell states with $j$ = $1/2$, $3/2$, $5/2$, and $7/2$ unambiguously.
In this coupling notation, an additional quantum number, $\alpha$, only 
occurs for subshell states with $j \ge 9/2$; for $j$ = $9/2$,
we use the quantum number $w = 0, \, 1,$ or $2$ similar as for $f-$shells in 
$LS-$coupling. A list of all allowed subshell states in $jj-$coupling with 
$j$ = $1/2$, $3/2$, $5/2$, $7/2$, and $9/2$ were given in Ref.\ 
\cite{Gaigalas:01}, table 2. In  fact, all $LS \leftrightarrow jj$
transformations of symmetry--adapted functions can always be traced back to 
the corresponding transformation of the subshell states 
(\ref{eq:quasispin-LS-a}) and (\ref{eq:quasispin-jj-a}), from which these
symmetry functions are built--up. 
 
\medskip 
 
Although, at the first glance, the definition of the subshell states in 
$LS-$ and $jj-$coupling appears very similar, these states generally belong 
to different irreducible representations of the $SO_3$ rotation group. 
In $jj-$coupling, each (nonrelativistic) $nl$--shell is usually 'separable' 
into two (relativistic) subshells with total angular momenta 
$j_{\pm} \,=\, l \pm 1/2 $. Therefore, in order to transform a shell state 
\ket{l^{\,N} \alpha L S} into a $jj-$coupled basis\footnote{Here and in 
the following, we often omit the principal quantum number $n$ in the notation 
of the subshell states as this quantum number is irrelevant for the 
transformation properties of these states. The principal quantum 
number is needed only if two or more subshell states with the same $l$ and 
$j$ but different $n$'s later occur in the construction of the symmetry 
functions. }, two subshell 
states with $j_{-}$ and $j_{+}$  may both occur in the expansion, i.e.\  
   \ket{\tmm{\kappa}^{\tmm{N}} \tmm{\nu} \tmm{J}} and
   \ket{\tpm{\kappa}^{\tpm{N}} \tpm{\nu} \tpm{J}},
where we utilize again the relativistic quantum number $\kappa$ to simplify 
the notation below. Obviously, also, $ N \,=\, \tmm{N} + \tpm{N} $ and  
$\tmm{\kappa} \,=\, -(\tpm{\kappa}+1) \,>\,0$ must hold where the notation 
$\,\tmm{\kappa} \;>\,0$ and $\,\tpm{\kappa} \;<\,0$ becomes clearer when 
one considers the corresponding total angular momentum 
$j_{\pm} \,=\, l \pm 1/2 $.  
 
\newpage
 
Making use of this notation, the transformation between the subshell states 
in $LS-$ and $jj-$coupling can be written as 
\footnotesize
\begin{eqnarray} 
\label{Matrixjj-LSDef-one-a} 
   \ketm{l^{\,N} \; \alpha LS\ J}  
   & = & 
   \sum_{\tmm{N} \tmm{\nu} \tmm{J} \tpm{\nu} \tpm{J}}   \, 
   \ketm{(\tmm{\kappa}^{\tmm{N}}   \tmm{\nu} \tmm{J},  \ 
          \tpm{\kappa}^{(N-\tmm{N})} \tpm{\nu} \tpm{J})\ J} \, 
   \sprm{(\tmm{\kappa}^{\tmm{N}}   \tmm{\nu} \tmm{J},  \ 
          \tpm{\kappa}^{(N-\tmm{N})} \tpm{\nu} \tpm{J})\ J}{ 
          l^{\,N} \; \alpha LS\ J} 
   \\[0.2cm] 
\label{Matrixjj-LSDef-one-b} 
   \ketm{(\tmm{\kappa}^{\tmm{N}} \tmm{\nu} \tmm{J},\ 
          \tpm{\kappa}^{\tpm{N}} \tpm{\nu} \tpm{J})\ J}  
   & = & 
   \sum_{\alpha LS} \, 
   \ketm{l^{\,(\tmm{N}+\tpm{N})} \; \alpha LS\ J} 
   \sprm{l^{\,(\tmm{N}+\tpm{N})} \; \alpha LS\ J}{ 
         (\tmm{\kappa}^{\tmm{N}} \tmm{\nu} \tmm{J},\ 
          \tpm{\kappa}^{\tpm{N}} \tpm{\nu} \tpm{J})\ J} 
\end{eqnarray} 
\normalsize 
which, in both cases, includes a summation over all the quantum numbers 
(except of $\kappa$ and $l$). Here, 
\ket{(\tmm{\kappa}^{\tmm{N}} \tmm{\nu} \tmm{J}, \ 
      \tpm{\kappa}^{\tpm{N}} \tpm{\nu} \tpm{J})\ J} 
is a coupled state with well--defined total angular momentum $J$ which is  
built from the corresponding $jj-$coupled subshell states with  
$j_{\pm} \,=\, l \pm 1/2 $ and the total subshell angular momenta $\tmm{J}$ 
and $\tpm{J}$, respectively. 
 
\bigskip 
 
An explicit expression for the transformation coefficients  
\small 
\begin{eqnarray} 
\label{subshell-coefficients}
   \sprm{(\tmm{\kappa}^{\tmm{N}}   \tmm{\nu} \tmm{J},  \ 
          \tpm{\kappa}^{(N-\tmm{N})} \tpm{\nu} \tpm{J})\ J}{ 
          l^{\,N} \; \alpha LS\ J} 
   \; = \; 
   \sprm{l^{\,(\tmm{N}+\tpm{N})} \; \alpha LS\ J}{ 
         (\tmm{\kappa}^{\tmm{N}} \tmm{\nu} \tmm{J},\ 
          \tpm{\kappa}^{\tpm{N}} \tpm{\nu} \tpm{J})\ J}   
\end{eqnarray} 
\normalsize 
in (\ref{Matrixjj-LSDef-one-a}) and (\ref{Matrixjj-LSDef-one-b}) can be 
obtained only if we take the construction of the subshell states of $N$ 
equivalent electrons from their corresponding \textit{parent states} with 
$N-1$ electrons into account. For a number of special configurations,
expressions for these coefficients have been displayed before in Ref.\ 
\cite{Gaigalas/ZR:01}. In
general, however, the \textit{recursive} definition of the subshell states,
out of their parent states, also leads to a recursive generation of the
transformation matrices (\ref{subshell-coefficients}) which we
summarize in Appendix A. For the moment, it is sufficient to say that
these transformation coefficients can be chosen \textit{real} and that 
they occur very frequently as the \textit{building blocks} in the 
transformation of all symmetry functions. The transformation 
matrices (\ref{subshell-coefficients}) are therefore implemented in a suitable 
form for all (sub--) shells with $l \le 3$ and occupation numbers 
$ N \,=\, 1,\ 2,\ ...,\ 2l+1$ in the current extension to the \textsc{Racah} 
program. For all other allowed occupation numbers $ N \,=\, 2l+2,\ ...,\ 4l+2$,
these transformation coefficients are obtained according to their
\textit{electron--hole} symmetry from the matrix elements for 
$ N' \,=\, 4l+2-N$. Such a symmetry relation was established originally
by Grant \textit{et al.} \cite{Dyall_Grant} and later utilized also in the
tabulations of Gaigalas \textit{et al.} \cite{Gaigalas/ZR:02} for all 
subshells with $l \leq 3$.

\subsection{Coupling of subshell states} 
 
Of course, many electron configurations with a single open shell occur in
the notation of atomic levels and may allow a rough characterization.
For a detailed representation of these levels, however, configuration state
functions (CSF) with several open shell need to be taken into account, 
a situation which is even strongly enhanced when open $d-$ or $f-$shell 
elements or excited levels are to be considered. In such situations, the 
construction of a suitable symmetry--adapted basis for the representation of 
the atomic states also requires the coupling of two or more (open) subshell 
states. In $LS-$coupling, typically, a CSF basis is constructed from a 
stepwise coupling of the individual shells  
$l_1^{\,N_1},\ l_2^{\,N_2},\ ...$ 
\begin{eqnarray} 
\label{LS-CSF} 
   \ketm{(...(((l_1^{\,N_1} \alpha_1 L_1 S_1, \,  
                l_2^{\,N_2} \alpha_2 L_2 S_2) L_{12}S_{12}, \, 
                l_3^{\,N_3} \alpha_3 L_3 S_3) L_{123}S_{123})...)\ J} 
\end{eqnarray} 
which could be written explicitly also in terms of a Clebsch--Gordan  
expansion. For the case of two open shells, for example, a proper CSF 
basis   
\begin{eqnarray} 
\label{two-shell-csf} 
   & & \hspace*{-1.5cm} 
   \ketm{(l_1^{\,N_1} \alpha_1 L_1 S_1,\, l_2^{\,N_2} \alpha_2 L_2 S_2) LS\, J} 
   \nonumber \\[0.2cm] 
   & = & 
   \sum_{M_{L_1} M_{S_1} M_{L_2} M_{S_2} M_L M_S} 
   \ketm{l_1^{N_1} \alpha_1 L_1 S_1 \, M_{L_1} M_{S_1}} 
   \ketm{l_2^{N_2} \alpha_2 L_2 S_2 \, M_{L_2} M_{S_2}} 
   \nonumber \\[0.2cm] 
   &  & \hspace*{1.0cm} \times 
   \sprm{L_{1} M_{L_1} \, L_{2} M_{L_2}}{L M_{L}} 
   \sprm{S_{1} M_{S_1} \, S_{2} M_{S_2}}{S M_{S}} 
   \sprm{L M_{L} \, S M_{S}}{J M_{J}} 
\end{eqnarray} 
includes the coupling of the subshell orbital angular momenta $L_1$ and $L_2$  
to a total $L$ and the subshell spins $S_1$ and $S_2$ to a total $S$ which are  
finally coupled to a total $J$. 
 
\medskip 
 
A very similar sequence for the coupling of the subshell states  
$ \ketm{\kappa_1^{\,N_1} \; \nu_1 J_1},\ 
  \ketm{\kappa_2^{\,N_2} \; \nu_2 J_2},\ ... $  
is applied also in $jj-$coupling. In principle, again, any (predefined)  
sequence of the $jj-$coupled subshells will give rise to a valid  
many--particle basis. 
For practical purposes and in particular for an efficient transformation of 
such configuration states, however, it is useful to define a 
\textit{standard order} for $jj-$coupled configuration states such as
\vspace*{-0.1cm}
 
\footnotesize 
\begin{eqnarray} 
\label{standard-order} 
   \ketm{(...(((((\tmm{\kappa}_1^{\tmm{N}_1} \tmm{\nu}_1 \tmm{J}_1,         \, 
             \tpm{\kappa}_1^{\tpm{N}_1} \tpm{\nu}_1 \tpm{J}_1) J_1,         \, 
             \tmm{\kappa}_2^{\tmm{N}_2} \tmm{\nu}_2 \tmm{J}_2) J^{\prime}_{12},\, 
             \tpm{\kappa}_2^{\tpm{N}_2} \tpm{\nu}_2 \tpm{J}_2) J_{12} 
             \tmm{\kappa}_3^{\tmm{N}_3} \tmm{\nu}_3 \tmm{J}_3) J^{\prime}_{123},\, 
             \tpm{\kappa}_3^{\tpm{N}_3} \tpm{\nu}_3 \tpm{J}_3) J_{123}...)J} 
\end{eqnarray} 
\normalsize 
which fulfills two additional conditions: 
 
(i) If both subshells with common $l_i$, i.e.\ $ \tmm{\kappa}_i$ and  
    $ \tpm{\kappa}_i$ appears in the expansion, these two subshells always 
    occur successively in the sequence 
    $ (\tmm{\kappa}_i^{\tmm{N}_i} \tmm{\nu}_i \tmm{J}_i, \, 
       \tpm{\kappa}_i^{\tpm{N}_i} \tpm{\nu}_i \tpm{J}_i) \ J_{i} $. 
    Formally, we can use this sequence even for subshell states 
    with zero occupation if we interprete  
    $\ketm{\kappa^0 \nu=0\ J=0} \,\equiv\, 1$; in this case, the full 
    Clebsch--Gordan expansion [cf.\ (\ref{two-shell-csf})] remains valid due  
    to the orthonormality properties of the Clebsch--Gordan coefficients. 
     
\smallskip 
     
(ii) For the $LS-jj$ transformation of configuration states 
\small 
\begin{eqnarray} 
\label{general-trans}
   & & \hspace*{-0.8cm} 
   \left\langle 
    (...(((l_1^{\,N_1} \alpha_1 L_1 S_1, \,  
           l_2^{\,N_2} \alpha_2 L_2 S_2) L_{12}S_{12}, \, 
           l_3^{\,N_3} \alpha_3 L_3 S_3) L_{123}S_{123})...)\ J     
   \left\vert \right. 
    (...(((((\tmm{\kappa}_1^{\tmm{N}_1} \tmm{\nu}_1 \tmm{J}_1,         \, 
             \tpm{\kappa}_1^{\tpm{N}_1} \tpm{\nu}_1 \tpm{J}_1) J_1, \right. 
   \nonumber \\[0.2cm]  
   & & \hspace*{3.0cm} 
   \left. 
             \tmm{\kappa}_2^{\tmm{N}_2} \tmm{\nu}_2 \tmm{J}_2)J^{\prime}_{12}, \, 
             \tpm{\kappa}_2^{\tpm{N}_2} \tpm{\nu}_2 \tpm{J}_2) J_{12} 
             \tmm{\kappa}_3^{\tmm{N}_3} \tmm{\nu}_3 \tmm{J}_3)J^{\prime}_{123},\, 
             \tpm{\kappa}_3^{\tpm{N}_3} \tpm{\nu}_3 \tpm{J}_3) J_{123}...)\ J 
   \right\rangle 
\end{eqnarray} 
\normalsize 
    we further assume in \textit{standard order} that 
    $ l_1 \,=\, \tmm{\kappa}_1 \,=\, -(\tpm{\kappa_1}+1),\   
      l_2 \,=\, \tmm{\kappa}_2 \,=\, -(\tpm{\kappa_2}+1),\  ...$, i.e.\ 
    that the sequence of (sub--)shell states is the 
    \textit{same on both sides} of the transformation matrix. 
     
\smallskip 
     
In the following, we always assume this standard order in the derivation of 
explicit expressions for the $LS-jj$ transformation matrices. The two 
conditions (i--ii) simplifies the implementation of these matrices 
considerably without much loss of generality. The same or at least a
very similar \textit{order} in the sequence of the individual shells is 
assumed in many standard programs on atomic structure. 
 
\medskip 
 
We are now prepared to write down the transformation coefficients  
\small 
\begin{eqnarray} 
\label{2-shell-matrix} 
   \sprm{(((\tmm{\kappa}_1^{\tmm{N}_1} \tmm{\nu}_1 \tmm{J}_1,         \, 
            \tpm{\kappa}_1^{\tpm{N}_1} \tpm{\nu}_1 \tpm{J}_1) J_1,    \, 
            \tmm{\kappa}_2^{\tmm{N}_2} \tmm{\nu}_2 \tmm{J}_2) J^{\prime}_{12},\, 
            \tpm{\kappa}_2^{\tpm{N}_2} \tpm{\nu}_2 \tpm{J}_2) \ J}{ 
         (l_1^{\,N_1} \alpha_1 L_1 S_1, \, 
          l_2^{\,N_2} \alpha_2 L_2 S_2) LS \, J}  
\end{eqnarray} 
\normalsize 
for the coupling of two open shells in $LS-$coupling or up to four open 
subshells in $jj-$coupling, respectively. As before, an expansion of a 
$LS-$coupled CSF in an appropriate  $jj-$coupled basis is obtained by a 
summation over all quantum numbers apart from the $\kappa_i$'s. Making use 
of the two expansions (\ref{Matrixjj-LSDef-one-a}), (\ref{two-shell-csf}) 
and of the recoupling theory of angular momenta, we easily find 
\small

\begin{eqnarray} 
\label{exapmle-b} 
   & & \hspace*{-0.8cm} 
   \ketm{(l_1^{\,N_1} \alpha_1 L_1 S_1, \, 
          l_2^{\,N_2} \alpha_2 L_2 S_2) LS \, J} 
   \nonumber \\[0.2cm] 
   & = & 
   \sum_{\hspace*{-0.5cm}
   \tmm{N}_1 \tmm{N}_2 \tmm{\nu}_1 \tmm{J}_1 \tpm{\nu}_1 \tpm{J}_1
                             \tmm{\nu}_2 \tmm{J}_2 \tpm{\nu}_2 \tpm{J}_2  
         J_1 J_2 J^{\prime}_{12}} \, 
   \ketm{(((\tmm{\kappa}_1^{\tmm{N}_1} \tmm{\nu}_1 \tmm{J}_1,               \, 
            \tpm{\kappa}_1^{(N_1-\tmm{N}_1)} \tpm{\nu}_1 \tpm{J}_1) J_1,    \, 
            \tmm{\kappa}_2^{\tmm{N}_2} \tmm{\nu}_2 \tmm{J}_2) J^{\prime}_{12}, \, 
            \tpm{\kappa}_2^{(N_2-\tmm{N}_2)} \tpm{\nu}_2 \tpm{J}_2) \ J }   \, 
   \nonumber \\[0.4cm] 
   &  & \hspace*{1.0cm} \times 
   \sprm{((L_1,L_2)L,(S_1,S_2)S)J}{((L_1,S_1)J_1,(L_2,S_2)J_2)J} \, 
   \nonumber \\[0.2cm] 
   &  & \hspace*{1.0cm} \times 
   \sprm{((\tmm{J_1},\tpm{J_1})J_{1},(\tmm{J_2},\tpm{J_2})J_2)J} 
   {(((\tmm{J_1},\tpm{J_1})J_{1},\tmm{J_2})J^{\prime}_{12},\tpm{J_2})J} \, 
   \nonumber \\[0.2cm] 
   &  & \hspace*{1.0cm} \times  
   \sprm{(\tmm{\kappa}_1^{\tmm{N}_1} \tmm{\nu}_1 \tmm{J}_1, \ 
          \tpm{\kappa}_1^{(N_1-\tmm{N}_1)} \tpm{\nu}_1 \tpm{J}_1) J_1}{ 
          l_1^{N_1} \; \alpha_1 L_1 S_1\ J_1}  \; 
   \nonumber \\[0.2cm] 
   &  & \hspace*{1.0cm} \times 
   \sprm{(\tmm{\kappa}_2^{\tmm{N}_2} \tmm{\nu}_2 \tmm{J}_2, \ 
          \tpm{\kappa}_2^{(N_2-\tmm{N}_2)} \tpm{\nu}_2 \tpm{J}_2) J_2}{ 
          l_2^{N_2} \; \alpha_2 L_2 S_2\ J_2}  \, 
\end{eqnarray}
\normalsize 
where the third and fourth line denote two \textit{recoupling coefficients} 
which ensure that the sequence of the couplings on the left-- and right--hand 
side of the expansion is taken into account properly. 
These recoupling coefficients can be evaluated 
by means of the \textsc{Racah} program; here we postpone the derivation 
until section 5 and display only the final result 
\small
\begin{eqnarray} 
\label{exapmle-explicit} 
   & & \hspace*{-0.8cm} 
   \ketm{(l_1^{\,N_1} \alpha_1 L_1 S_1, \, 
          l_2^{\,N_2} \alpha_2 L_2 S_2) LS \, J} 
   \nonumber \\[0.2cm] 
   & = & 
   \sum_{\hspace*{-0.3cm}
         \tmm{N}_1 \tmm{N}_2 \tmm{\nu}_1 \tmm{J}_1 \tpm{\nu}_1 \tpm{J}_1 
                             \tmm{\nu}_2 \tmm{J}_2 \tpm{\nu}_2 \tpm{J}_2  
         J_1 J^{\prime}_{12} } \, 
   \ketm{(((\tmm{\kappa}_1^{\tmm{N}_1} \tmm{\nu}_1 \tmm{J}_1,               \, 
            \tpm{\kappa}_1^{(N_1-\tmm{N}_1)} \tpm{\nu}_1 \tpm{J}_1) J_1,    \, 
            \tmm{\kappa}_2^{\tmm{N}_2} \tmm{\nu}_2 \tmm{J}_2) J^{\prime}_{12}, \, 
            \tpm{\kappa}_2^{(N_2-\tpm{N}_2)} \tpm{\nu}_2 \tpm{J}_2) J }     \, 
   \nonumber \\[0.4cm] 
   &   & \hspace*{1.0cm} \times \;
   (-1)^{\tmm{J}_2+\tpm{J}_2+J_1+J} \; 
   \sqrt{\jfactorm{J_1,J^{\prime}_{12},L,S}} \;
   \sprm{(\tmm{\kappa}_1^{\tmm{N}_1} \tmm{\nu}_1 \tmm{J}_1, \ 
          \tpm{\kappa}_1^{\tpm{N}_1} \tpm{\nu}_1 \tpm{J}_1)\ J_1}{ 
          l_1^{\,N_1} \; \alpha_1 L_1 S_1\ J_1}  \, 
   \nonumber \\[0.2cm] 
   &   & \hspace*{1.0cm} \times \; \;  
   \sum_{J_2} 
   \jfactorm{J_2} \, 
   \ninejm{L_1}{S_1}{J_1}{L_2}{S_2}{J_2}{L}{S}{J} \, 
   \sixjm{J_1}{\tmm{J}_2}{J^{\prime}_{12}}{\tpm{J}_2}{J}{J_2} \;
   \nonumber \\[0.2cm] 
   &   & \hspace*{1.0cm} \times \;  
   \sprm{(\tmm{\kappa}_2^{\tmm{N}_2} \tmm{\nu}_2 \tmm{J}_2, \ 
          \tpm{\kappa}_2^{\tpm{N}_2} \tpm{\nu}_2 \tpm{J}_2) J_2}{ 
          l_2^{N_2} \; \alpha_2 L_2 S_2\ J_2}  \, 
\end{eqnarray} 
\normalsize
 
Explicit representations of the transformation matrices becomes quickly 
cumbersome if more than two open shells are involved. For such complex 
shell structures, a symbolic and automatic treatment seems to be inevitable.
In the present implementation of the \textsc{Racah} program, an automatic
transformation of configuration state functions is also restricted to 
symmetry functions with two open ($LS$--) shells, see section 4.

\subsection{Transformation of atomic states} 
 
Any set of configuration state functions (\ref{LS-CSF}) or  
(\ref{standard-order}) can be utilized to form a many--particle basis for 
the set--up of the (corresponding) Hamiltonian matrix. 
The diagonalization of this matrix then results in approximate atomic states, 
given as a superposition of the accordings CSF. While $LS-$coupled 
CSF (\ref{LS-CSF}) are frequently applied in nonrelativistic atomic structure 
calculations, the $jj-$coupled CSF (\ref{standard-order}) are the basis of 
most relativistic computations. However, not much need to be said here 
about the details of atomic structure theory. For the sake of simplicity, 
we can restrict ourselves to the configuration interaction (CI) approach 
in which (any) approximate atomic state is either written in terms of a 
$LS-$coupled  
\begin{eqnarray} 
\label{Psi-LS} 
   \ketm{\Psi_\alpha (J^P)} & = & 
   \sum_{r} \, a_r^{\,(LS)} (\alpha) \, \ketm{\gamma_r\ LS\ JP} 
\end{eqnarray} 
or $jj-$coupled basis 
\begin{eqnarray} 
\label{Psi-jj} 
   \ketm{\Psi_\alpha (J^P)} & = & 
   \sum_{s} \, a_s^{\,(jj)} (\alpha) \, \ketm{\gamma_s JP} \; ,
\end{eqnarray} 
where $\alpha \,=\, 1,\ 2,\ ... $ enumerate the atomic levels (of the given 
symmetry) and $\gamma_r,\ \gamma_s$ denote the sets of all quantum numbers  
as required for an unique classification of the CSF. No further information  
is needed to understand the transformation behaviour when one wants to 
transform a state \ket{\Psi_\alpha (J^P)} from one into another basis:
$$ \ketm{\Psi_\alpha^{\,(LS)} (J^P)}  \;\longleftrightarrow\;  
   \ketm{\Psi_\alpha^{\,(jj)} (J^P)} \, .$$ 
Obviously, the transformation of any atomic state can be reduced simply to 
the transformation of the underlying configuration state functions  
\ket{\gamma_r\ LS\ JP} and \ket{\gamma_s JP}, respectively. For these CSF, 
explicit expressions have been discussed in the previous subsection. 
 
\medskip 
 
From the user's viewpoint, of course, the main request to the present
implementation concerns the transformation of atomic states 
between a $LS-$ and $jj-$coupled basis. For these two coupling schemes, both 
directions of the transformation are equally supported, including a $LS-$ 
and $SL-$coupling sequence. 
To facilitate the use of the program as well as the communication inside 
of the code, a powerful notation has been introduced to define and to 
manipulate atomic states of type (\ref{Psi-LS}) or (\ref{Psi-jj}); 
see section 4 and Appendix B.1 below. With this additional notations, the 
transformation of atomic states is achieved by a single line which can be
typed in interactively. But already from the notation in this section
it becomes clear how (easily) other coupling schemes such as $JK-$ or
$LK-$ coupling \cite{Martin/Wiese:98}
could be incorporated into the program, for instance, 
in order to search for the \textit{best} representation of some level(s) 
in terms of a single coupling scheme.

%
%
%
%
 
\begin{table} 
\begin{small} 
 
{\bf Table 1} 
\hspace{0.2cm} 
{\rm  
Additional commands to the \textsc{Racah} package for $LS-jj$  
transformations of coupled (sub--) shell states. A more detailed description 
of these new procedures, which are provided for interactive work, is given 
in Appendix B. Our new commands also include a number of auxiliarity 
procedures to facilitate the \textit{communication} with and within the 
\textsc{Racah} program. 
} 
\begin{center} 
\begin{tabular}{l c p{10.6cm}} \\[-0.3cm]   \hline \hline   \\[-0.3cm] 
   shell\_{}jj(), shell\_{}LS()   & &  
      To represent a (sub--) shell state in either $jj-$ or  
      $LS-$coupling.                                                  \\[0.1cm] 
   csf\_{}jj(), csf\_{}LS()       & &  
      To represent a configuration state functions of one or two 
      open shells in either $jj-$ or $LS-$coupling.                   \\[0.1cm] 
   Racah\_{}matrix\_{}LS\_{}jj()  & & 
      Returns the $LS-jj$ transformation matrix \newline
      \spr{l^{\,N} \nu L S\ J}{(\tmm{\kappa}^{\,\tmm{N}} \tmm{\nu} \tmm{J},\, 
                                \tpm{\kappa}^{\,\tpm{N}} \tpm{\nu} \tpm{J} )\ J} 
      for all subshells with $l \, = \, 0,\,\ldots,\,3$.          \\[0.1cm] 
   Racah\_{}transform\_{}csf()    & & 
      Transforms a configuration state function from $jj-$ to  
      $LS-$ coup\-ling or vice versa. CSF of different 
      complexity are supported by the program.                        \\[0.1cm] 
   Racah\_{}transform\_{}asf()    & & Transforms an atomic state function 
      (ASF) from $jj-$ to $LS-$coupling or vice versa.                \\[0.1cm] 
   Racah\_{}set\_{}coupling()     & & \textit{Defines} the coupling order 
      in the set--up of the $LS-$ shell states to either $LS-$ or 
      $SL-$coupling.                                                  \\[0.1cm] 
   \hline \hline  
\end{tabular} 
\end{center} 
\end{small} 
 
\vspace{0.7cm} 
\end{table}

\section{New procedures for the \textsc{Racah} package} 
 
Since its first publication in 1997, the \textsc{Racah} package  
\cite{Fritzsche:97} has grown considerably. In its early days, emphasis was 
put on simplifying those Racah expressions which included summations only over 
the Wigner $n-j$ symbols [cf.\ Ref.\ \cite{Fritzsche:97}, Figure 1]. Apart 
from such \textit{algebraic} manipulations of expressions, however, recent 
developments also concerned an improved \textit{numerical} support of a few 
standard tasks from the theory of angular momentum. Following this line, 
a number of basic quantities for evaluating (many--electron) matrix elements 
have been provided recently \cite{Gaigalas:01} and will be further enhanced in
the future. With the present extension to the \textsc{Racah} program, we 
now facilitate also the access to $LS-jj$ transformation matrices and to the 
transformation of general atomic and configuration symmetry functions at 
different level of complexity.

\medskip 
 
Obviously, any automatic transformation of atomic states must enable the user
with a quick and simple access to the underlying symmetry functions. 
To support the \textit{definition} and communication of such functions,
two basic \textit{terms} from the atomic shell model play a central role:
atomic \textit{shell states} and their successive coupling which, finally, 
leads to the definition of \textit{configuration state functions (CSF)}. 
In order to simplify the handling of these functions, four \textit{auxiliarity} 
procedures have been designed and introduced into the \textsc{Racah} 
program with the intention to keep all necessary information about a subhsell
state or a CSF close together. These procedures such as \,\procref{csf\_{}jj()},
\procref{shell\_{}jj()}, \ldots\, are defined separately for each coupling 
scheme, i.e.\ for $LS-$ or $jj-$coupling, and, hence, could be easily 
extended to include other coupling schemes in the future. 
The procedure \procref{Racah\_{}set\_{}coupling()} also 
differs from most other commands in that it just 'assigns' the 
\textit{coupling order} $LS$ or $SL$ of the orbital angular momenta and the 
spins to a \textit{global} variable; this particular procedure must therefore
be invoked \textit{prior} to any other command which deals with  $LS-jj$ 
transformations. 
 
\medskip 
 
Although a simple notation has been worked out in order to construct CSF of 
any complexity, an automatic transformation of atomic states is currently
supported only for configuration states with up to two open shells in $LS-$ 
and up to four open subshells in $jj-$coupling, respectively. This limitation 
has arose from the number of recoupling coefficients which occur in any 
transformation and which grows rapidly if more open shells get involved. 
In the future, the program might be extended to more complex shell structures 
if demands arise from our work or from the side of users.
 
\medskip 
 
At user's level, only a very few procedures have to be known in order to 
obtain the $LS-jj$ transformation matrices or to transform some atomic or 
configuration states. Table 1 displays a short list of the main commands.
Owing to the rapid increase of the complexity of many expressions, however, 
a (much) larger number of procedures had to be implemented at a lower level 
of the program. Many of them make use of previous development such as the
routines for obtaining reduced coefficients of fractional parentage or for
evaluating recoupling matrices. Since all transformation coefficients are 
evaluated directly to their numerical (i.e.\ either algebraic or 
floating--point) values, no additional data structures needed to be defined 
for the present work. 
 
\medskip 
 
The explicit expression (\ref{exapmle-explicit}) of a transformation matrix 
for two open shells showed that these coefficients are always reduced to the
transformation of individual subshell states
$$ \sprm{l^{N} \; \alpha LS\ J}{ 
         \left( \tmm{\kappa}^{\tmm{N}} \tmm{\nu} \tmm{J}, \ 
                \tpm{\kappa}^{\tpm{N}} \tpm{\nu} \tpm{J} \right) J}  \; .$$
These matrix elements are stored internally in the program using the format
\begin{center} 
\texttt{[Q, L, S, J\_{}big, N\_{}1, Q\_{}1, J\_{}big\_{}1, Q\_{}2, 
J\_{}big\_{}2, factor, nom, denom]} 
\end{center} 
for l = 1, 2 and 
\begin{center} 
\texttt{[w, Q, L, S, J\_{}big, N\_{}1, Q\_{}1, J\_{}big\_{}1, 
Q\_{}2, J\_{}big\_{}2, factor, nom, denom]} 
\end{center} 
for $l = 3$, as their recursive computation (from the matrix elements of the
corresponding parents states) was found too slow for practical purposes. 
In this representation, the value of each $LS-jj$ transformation matrix 
elements is given by $factor \times  \sqrt{\frac{nom}{denom}}$;
they are kept for all occupation numbers $N \,=\, 1,\ ...,\ 2l+1$. 
 
\medskip 

With the increased size of the \textsc{Racah} program, a new \textit{program
structure} became necessary. Following recent suggestions by the \textsc{Maple} 
standard, therefore, the program is now divided into the two modules 
\texttt{Racah} and \texttt{Jucys} which must be loaded separately by the
\texttt{with()} feature of \textsc{Maple}. While the \texttt{Racah} module now 
contains all procedures for the set--up and manipulation of Racah expression
\cite{Fritzsche:97,Inghoff/Fri:01}, the standard quantities
from Ref.\ \cite{Gaigalas:01} and the present implementation of the $LS-jj$
transformation matrices are incorporated into the module \texttt{Jucys}.
Of course, the use of modules also helps to keep all low--level procedures
invisible to the user. The \textsc{Racah} package is distributed as a tar
(-xvf) file \texttt{Racah2002.tar} which, apart of the source code and
module libraries in different \textsc{Maple} versions, includes a 
\texttt{Read.me} for the installation of the program as well as the document 
\texttt{Racah-commands.ps}. This document provides the definition of all
\textit{data structures} of the \textsc{Racah} program as well as an 
alphabetic list of all user relevant commands. The code can be down--loaded also
from our home page via the world--wide--web 
(http://www.physik.uni-kassel/fritzsche). For most commands, moreover,
there are on--line help pages available which are distributed and maintained together
with the code.

\section{Examples} 
 
A few examples from atomic shell theory are displayed below to illustrate 
the application of the $LS-jj$ transformation matrices from section 2. 
Beside of the computation of a particular matrix elements, we briefly 
explain how, for instance, the user can generate a (full) $LS-jj$ 
transformation matrix for a half--filled $f-$shell, i.e.\ the subspace which 
is spanned by the \ket{f^{\,7}\ w \nu LS} subshell states. We also show 
explicitly how, for C$^{2+}$ ions, the two low--lying $ 1s^2 2s2p \;\, J=1$ 
levels from a ($jj-$coupled) multiconfiguration Dirac--Fock calculation can 
be easily transformed into a more appropriate $LS-$coupled basis. 
 
\medskip 
 
Let us start with the (numerical) evaluation of the $LS-jj$ transformation  
matrix element\footnote{The capital $J$ on the left--hand side denotes the
spectroscopic notation for the orbital angular momentum $L = 7$ and
should not be confused with the total angular momentum of this shell 
state.}
$$ \sprm{f^{\,3}\ (w=1,\nu=3)\ ^2 J_{15/2}}{
         f_{7/2}^{\,3}\ (\tpm{\nu}=3)\ 15/2} $$ 
for a partially filled $f-$shell, i.e.\ for the two (sub--) shell states  
with quantum numbers 
$ l = 3,$ $ N = 3, \, w = 1, \, \nu = 3, \, L = 7, \, S = 1/2 $, and 
$ \tpm{j} = 7/2, \, \tpm{N} = 3, \, \tpm{\nu} = 3, \, \tpm{J} = 15/2$, 
respectively. Since $\tmm{N} = 0$ in this case, no  
\ket{f_{5/2}^{\,0} \tmm{\nu} \tmm{J}} subshell state occurs in the notation.  
For these quantum numbers, we obtain the transformation coefficient simply  
by typing 
\vspace{-0.2cm} 
\begin{flushleft} 
{\tt 
$ > $ Racah\_set\_{}coupling(LS);        \\ 
$ > $ T := Racah\_{}matrix\_{}LS\_{}jj(shell\_{}LS(3,3,1,3,7,1/2), 
           \hspace*{8.5cm}shell\_{}jj(-4,3,3,15/2));   \\[-0.2cm]  
} 
\end{flushleft} 
\begin{center} 
   T :=  .586845597~. 
\end{center} 
As previously, the same result can be obtained also in algebraic or 
prime--number representations if the proper keywords \textit{algebraic} or 
\textit{prime} are added to the parameter list. In prime--number notation, 
for instance, the transformation coefficient  
\vspace{-0.2cm} 
\begin{flushleft} 
{\tt 
$ > $ T := Racah\_{}matrix\_{}LS\_{}jj(shell\_{}LS(3,3,1,3,7,1/2), 
           \hspace*{8.5cm}shell\_{}jj(-4,3,3,15/2),\textit{prime}); \\[-0.2cm]  
} 
\end{flushleft} 
\begin{center} 
   T :=  [1, -3, 3, 1, -2] 
\end{center} 
is returned as a list of (the first few non--zero) integer powers  
\begin{eqnarray} 
\left[ 
a_{0},~a_{1},~a_{2},~a_{3},~a_{4},~a_{5},~a_{6},~a_{7},~ 
a_{8},~a_{9},~a_{10},~a_{11} 
\right] . 
\nonumber 
\end{eqnarray} 
of the prime numbers  
$p_1 = 2$, $p_2 = 3$, $p_3 = 5$, $p_4 = 7$, $p_5 = 11$, $p_6 = 13$,  
$p_7 = 17$, $p_8 = 19$, $p_9 = 23$, $p_{10} = 29$, $p_{11} = 31$ 
from which the actual values is obtained as 
\begin{eqnarray} 
\label{eq:nk} 
   a_{0} \, \left( \, \displaystyle {\stackrel{11}{\prod_{i=1}}} 
   p_{i}^{\,a_{i}} \, \right)^{\frac 12} \; . 
\end{eqnarray} 
That is, the result [1, -3, 3, 1, -2] is just equivalent to the value 
$ \frac{3}{2 \times 7} \: \sqrt{\frac{3 \times 5}{2}} \,\approx\, 0.586845597$. 
 
\bigskip 
 
Several tabulations have been published over the years  
\cite{Gaigalas/ZR:02,Calvert,Childs:97} to the $LS-jj$ transformation matrices  
\spr{l^{N} \; \alpha LS\ J}{(\tmm{\kappa}^{\tmm{N}} \tmm{\nu} \tmm{J}, \ 
                             \tpm{\kappa}^{\tpm{N}} \tpm{\nu} \tpm{J}) J}  
for different --- partially filled --- shells and using different phase
conventions. With the present extension to the \textsc{Racah} package, we now 
provide a much simpler access to these transformation matrices which can be 
adopted to the actual requirements of the user. As a second example, 
therefore, we display how one can \textit{create} an 'electronic table' 
(for any shell with $l \le 3$) within only a few lines of \textsc{Maple} code. 
Figure 1 shows the necessary code for a half--filled  $f^{\,7}$ shell which 
can easily be modified and extended for other shells as well. The printout 
from this example is shown below in the \textsc{Test Run Output}. 

\begin{figure} 
\vspace*{-0.6cm}
\begin{scriptsize} 
\begin{verbatim} 
   Racah_set_coupling(LS);                                  Racah_set_coupling_scheme(LS_quasispin); 
   l := 3;     N := 7;     j1 := 5/2;     j2 := 7/2;        t1 := Racah_subshell_term_LS(l,Q_int); 
   kappa_1 := 3;     kappa_2 := -4;                         Racah_set_coupling_scheme(jj_quasispin); 
   for J from 1/2 to (N*l)+1/2 do                           lprint("J=",J);   
     for i from 1 to nops(t1) do 
       if abs(t1[i][4]-t1[i][5])<=J and J<=t1[i][4]-t1[i][5] and type(J+t1[i][4]+t1[i][5],integer) then 
         s1 := shell_LS(l, N, t1[i][2], 2*l+1-2*t1[i][3], t1[i][4], t1[i][5], check); 
         lprint(Racah_shell_print(s1)); 
         for N1 from 0 to N do 
           M1 := (N1-(2*j1+1)/2)/2;     N2 := N-N1;     M2 := (N2-(2*j2+1)/2)/2; 
           if type(M1,integer) then 
             t2 := Racah_subshell_term_jj(j1,Q_int);        t3 := Racah_subshell_term_jj(j2,Q_int); 
           elif not type(M1,integer) then 
             t2 := Racah_subshell_term_jj(j1,Q_halfint);    t3 := Racah_subshell_term_jj(j2,Q_halfint); 
           fi; 
           for i_1 from 1 to nops(t2) do 
             if abs(M1) <= t2[i_1][3] then 
               s2 := shell_jj(kappa_1, N1, (2*j1+1)/2-2*t2[i_1][3], t2[i_1][4]); 
               for i_2 from 1 to nops(t3) do 
                 if abs(M2)  <=  t3[i_2][3]  then 
                  if abs(t2[i_1][4] - t3[i_2][4]) <= J  and  J <= t2[i_1][4] + t3[i_2][4]  then 
                     s3 := shell_jj(kappa_2, N2, (2*j2+1)/2-2*t3[i_2][3], t3[i_2][4], check); 
                     result := Racah_matrix_LS_jj(s1,s2,s3,J,prime); 
                     lprint(Racah_shell_print(s2),Racah_shell_print(s3),result); 
                 end if;  end if;  
   end do;  end if;  end do;  end do;  end if;  end do;  end do; 
\end{verbatim} 
\end{scriptsize} 
{\bf Figure 1:} 
\hspace{0.2cm} 
{\rm Generation of the $LS-jj$ transformation matrix for a half--filled 
$f^{\,7}$ shell. The beginning of this table is shown in the  
\textsc{Test Run Output} below.} 
 
 
\end{figure} 
 
\bigskip

We now extent our examples to the transformation of configuration state 
functions or even atomic states as they frequently appear in standard  
(relativistic) computations. To deal with a simple case, let us consider  
the two lowest $1s^2 2s 2p \;\, J = 1$ levels of C$^{2+}$ ions which  
attracted a lot of recent interest in the diagnostics of  
stellar atmospheres \cite{LeTeuff:00},
In a single--configuration approximation, these two levels are  
written in terms of only two $jj-$coupled CSF 
\begin{eqnarray} 
\label{CIII-expansion} 
    \ketm{\Psi_{\alpha}} & = & a_1 (\alpha) \,\ketm{\gamma_1\ J=1} \;+\; 
                               a_2 (\alpha) \,\ketm{\gamma_2\ J=1} 
\end{eqnarray} 
where $\alpha \,=\, 1,2$, and the configuration states 
\begin{eqnarray} 
   \ketm{\gamma_1\ J=1} & = &  
   \ketm{\left( 2s_{1/2}^1\ (\nu=1)\ 1/2;\ 
                2p_{1/2}^1\ (\nu=1)\,1/2 \right)\ J=1} 
   \nonumber \\[-0.1cm] 
\label{J-1-levels} 
   \\[-0.4cm] 
   \ketm{\gamma_2\ J=1} & = &  
   \ketm{(2s_{1/2}^1\ (\nu=1)\ 1/2;\ 2p_{3/2}^1\ (\nu=1)\,3/2)\ J=1} 
   \nonumber  
\end{eqnarray} 
are derived from the coupling of the two valence electrons in an open $s-$ and  
$p-$shell, respectively. In this notation, we omit the $1s^2$  
core since it does not \textit{take part} in the coupling of the shells 
or in according transformations. The mixing coefficients  
$\{ a_i\ (\alpha) \}$ in Eq.\ (\ref{CIII-expansion}) can be obtained from  
either a multiconfiguration Dirac--Fock (MCDF) or configuration interaction 
calculation. By using,  
for instance, the well--known \textsc{Grasp92} code \cite{Grasp92}, we  
find for the two $J=1$ levels the expansions 
\begin{eqnarray} 
    \ketm{\Psi_{1}} & = &  0.8170 \,\ketm{\gamma_1\ J=1} \;+\; 
                           0.5767 \,\ketm{\gamma_2\ J=1} 
    \nonumber \\[0.1cm] 
    \ketm{\Psi_{2}} & = & -0.5767 \,\ketm{\gamma_1\ J=1} \;+\; 
                           0.8170 \,\ketm{\gamma_2\ J=1} \; . 
\end{eqnarray} 
which clearly illustrate that a (pure) $jj-$coupling scheme is 
inappropriate for the present example. 
 
\medskip 
 
A (much) more appropriate representation might be obtained in $LS-$coupling. 
However, before we transform the two atomic states \ket{\Psi_{1,2}} 
into such a representation, we first show the transformation of a single
configuration state, say, \ket{\gamma_1\ J=1}. The quantum numbers of this 
state can be read off directly from its definition in (\ref{J-1-levels}). 
In the \textsc{Racah} program, we may enter this CSF as 
\begin{flushleft} 
{\tt 
$ > $ CSF\_{}1 := csf\_{}jj(shell\_{}jj(-1,1,1,1/2),shell\_{}jj(1,1,1,1/2),1,check);     
\\[-0.2cm]  
} 
\end{flushleft} 
\begin{verbatim} 
          CSF_l = csf_jj(shell_jj(-1,1,1,1/2),shell_jj(1,1,1,1/2),1) 
\end{verbatim} 
where use is made of the two auxiliarity procedures \texttt{csf\_jj()}  
and \texttt{shell\_jj()}. These procedures return the input (basically)  
\textit{unevaluated} but help facilitate the communication with and within  
the program (see Appendix B for further details about these commands). 
By \textit{defining} first (again) the coupling sequence for the $LS-$subshell 
states, we obtain the (complete) expansion of the CSF  \ket{\gamma_1\ J=1} 
by  
\begin{flushleft} 
{\tt 
$ > $ Racah\_set\_coupling(\textit{LS});                  \\ 
$ > $ Racah\_transform\_{}csf("jj-$>$LS",CSF\_1,\textit{print}):      
\\[-0.2cm]  
} 
\end{flushleft} 
\begin{verbatim} 
          ".5773502693 * |(s^1 nu=1, ^2S;  p^1 nu=1, ^2P) ^2P_1 >"
          ".8164965809 * |(s^1 nu=1, ^2S;  p^1 nu=1, ^2P) ^3P_1 >" 
\end{verbatim} 
where, in spectroscopic notation, the first lines represents the $^1P_1$
($L=1,\ S=0,\ J=1$) and the second the $^3P_1$ component.
In the last input line, the keyword \textit{print} causes the procedure 
to 'print' the result (and to return a \textsc{null} expression) while, 
otherwise, the same result is returned in a list structure  
[csf$_1$(), c$_1$, csf$_2$(), c$_2$, ...], as suitable for further 
manipulations. 
 
\medskip 
 
Having a simple access to the transformation of configuration states, 
we are now prepared to transform also the (full) atomic states.  
For the wave functions \ket{\Psi_{1,2}} of the two $J=1$ levels, this  
is simply achieved (as before) by assigning the CSF \ket{\gamma_2\ J=1}  
from  (\ref{J-1-levels}) also to some variable \texttt{CSF\_{}2}  
and by carrying out the transformation explicitly: 
\begin{flushleft} 
{\tt 
$ > $ CSF\_{}2 := csf\_jj(shell\_{}jj(-1,1,1,1/2),shell\_{}jj(-2,1,1,3/2),1); \\
$ > $ Racah\_transform\_{}asf("jj-$>$LS",CSF\_1,0.8170,CSF\_2,0.5767,
                              \textit{print}):
\\[-0.2cm]  
} 
\end{flushleft} 
\begin{verbatim} 
          ".9426 * |(s^1 nu=1, ^2S;  p^1 nu=1, ^2P) ^1P_1 >"
          ".3341 * |(s^1 nu=1, ^2S;  p^1 nu=1, ^2P) ^3P_1 >"
\end{verbatim} 
and
\begin{flushleft} 
{\tt 
$ > $ Racah\_transform\_{}asf("jj-$>$LS",CSF\_1,-0.5767,CSF\_2,0.8170,
                              \textit{print}):
\\[-0.2cm]  
} 
\end{flushleft} 
\begin{verbatim} 
         " .3341 * |(s^1 nu=1, ^2S;  p^1 nu=1, ^2P) ^1P_1 >"
         "-.9426 * |(s^1 nu=1, ^2S;  p^1 nu=1, ^2P) ^3P_1 >"  .  
\end{verbatim} 
Apparently, while the ground state level \ket{\Psi_{1}} is a $^1P_1$
level, \ket{\Psi_{2}} represents the $^3P_1$ level. Again, the \textit{print} 
flag is used and ensures that the results are printed to screen and are 
not returned in terms of a list structure.  
 
\medskip 
 
The command \procref{Racah\_transform\_{}asf()} provides a very flexible 
access to the transformation of atomic states; any number of CSF along  
with their corresponding mixing coefficients $a_i$ can appear in the 
parameter list. Moreover, the same syntax applies for this procedure if 
an atomic state is given in $LS-$coupling,
\begin{eqnarray} 
    \ketm{\Phi_{\alpha}} & = &  
    \sum_{i} \, a_i (\alpha) \,\ketm{\gamma_i\ LSJ} \, , 
\end{eqnarray} 
and should be transformed into $jj-$coupling. With just two minor 
differences: The string \texttt{"LS-$>$jj"} has to be used, instead, 
and the \texttt{CSF$_i$} in the  
parameter list must represent proper $LS-$coupled configuration state 
functions. Although the program is currently limited to two open  
shells, of course, the same \textit{syntax} could be used for more complex  
shell structures or if other coupling schemes are to be incorporated  into 
the program. Moreover, the construction of the CSF from the successive coupling 
of subshell states will help  tackle more enhanced tasks in the future, such
as the computation of \textit{angular coefficients} for (non--scalar)
tensorial operators of rank $K$. 

\section{Evaluation of transformation matrices} 

We now return to the evaluation of the transformation coefficients 
(\ref{general-trans}) from section 2 which we could easily write down
in this \textit{bra--ket} notation. However, to carry out any transformation 
of configuration or atomic states explicitly, these matrix elements 
must be simplified to a computationally suitable form. This is achieved by
the \textit{recoupling of the angular momenta} which, in a number of steps, 
enables us to bring them into an equal sequence on the 
left-- and right--hand side of the transformation matrix (\ref{general-trans}).
Therefore, any transformation matrix can always be expressed in terms of several 
recoupling coefficients and an appropriate number of transformation matrices
(\ref{subshell-coefficients}), i.e.\ 
$$\sprm{l^{N} \; \alpha LS\ J}{ 
         \left( \tmm{\kappa}^{\tmm{N}} \tmm{\nu} \tmm{J}, \ 
                \tpm{\kappa}^{\tpm{N}} \tpm{\nu} \tpm{J} \right) J} \; ,   $$  
where a single matrix occurs for each open shell in the construction 
of the symmetry--adapted functions.

\medskip

For $LS-jj$ transformations and the standard order (\ref{standard-order}) of 
the $jj-$coupled subshell states, \textit{two steps} are required for the 
recoupling of the angular momenta. The first step (i) arise from the 
recoupling of the total subshell orbital angular momenta $L_i$ and 
spins $S_i$ 
\small 
\begin{eqnarray} 
\label{general-rec-1} 
   &  &  \hspace*{-2.0cm}
   \left\langle 
    ((((L_1 , \, L_2) L_{12} , \, L_3) L_{123} , \, ... )L , \, 
    (((S_1 , \, S_2) S_{12} , \, S_3) S_{123} , \, ... )S)\ J  \right. 
   \left\vert  \right. 
    \nonumber \\[0.2cm]  &  &  \hspace*{3.0cm} \left. 
    ((((L_1 , \, S_1) J_{1} , \, (L_2 , \, S_2) J_{2} )J_{12} , 
    (L_3 , \, S_3) J_{3} )\ J_{123} 
     ...)\ J 
   \right\rangle \; ,
\end{eqnarray} 
\normalsize 
in order to obtain the total subshell angular momenta $J_i$. In the second 
step, then, (ii) these total angular momenta are brought into their standard
order (\ref{standard-order}) by
\small 
\begin{eqnarray} 
\label{general-rec-2} 
   &  &  \hspace*{-1.0cm}
   \left\langle 
    ((((\tmm{J}_1 , \, \tpm{J}_1) J_{1} ,\, (\tmm{J}_2 ,\, 
                                             \tpm{J}_2) J_{2})J_{12} , \, 
       (\tmm{J}_3 , \, \tpm{J}_3) J_{3}  \, ) J_{123} , \, ... \, ) \, \ 
                                               J  \right. 
    \vert  
    \nonumber \\[0.02cm]  &  &  \hspace*{3.0cm} \left. 
    ((((\tmm{J}_1 , \, \tpm{J}_1) J_{1} ,\, \tmm{J}_2 ,\, ) 
                                         J^{\prime}_{12} ,\, 
    \tpm{J}_2) J_{12} ,  
    \tmm{J}_3 \, ) J^{\prime}_{123} , \, 
    \tpm{J}_3) J_{123} , \, ... \, ) \, \ J  
    \right\rangle \, . 
\end{eqnarray} 
\normalsize 
Further steps in the recoupling of angular momenta may arise if the
$jj-$coupled CSF are not defined in standard order (\ref{standard-order})
or if more elaborate coupling schemes occur. 

\medskip

Recoupling coefficients of type (\ref{general-rec-1}) and 
(\ref{general-rec-2}) can be evaluated by means of the \textsc{Racah} program
\cite{Fri/Inghoff:01}. In the following, we demonstrate this recent progress
with the simplification of these coefficients for configuration states with 
two open shells in $LS-$coupling \newline
$\ketm{(l_1^{N_1} \alpha_1 L_1 S_1, \, l_2^{N_2} \alpha_2 L_2 S_2) LS \, J}$. 
In this case, the first recoupling coefficient (\ref{general-rec-1})
simplifies to \spr{((L_1,L_2)L,(S_1,S_2)S)J}{((L_1,S_1)J_1,(L_2,S_2)J_2)J} 
and is evaluated  interactively by
\begin{flushleft} 
{\tt 
$ > $ rcc\_1 := Racah\_set(recoupling(`<((L1,L2)L,(S1,S2)S)J| \\
                \hspace*{8.8cm} ((L1,S1)J1,(L2,S2)J2)J>`)): \\ 
$ > $ rcc\_1 := Racah\_evaluate(rcc\_1):  \\
$ > $ Racah\_print(rcc\_1): 
\\[-0.2cm]  
}
\newpage
\end{flushleft} 
\vspace{-0.2cm} 
\begin{center} 
\begin{verbatim} 
--->
                     (-2 Ll + 2 S2 + 2 S1 + 2 L2 + 2 J)
                 (-1)
\end{verbatim} 
$\sqrt{{\tt 2 J2 + 1}}\sqrt{{\tt 2 J1 + 1}}\sqrt{{\tt 2 S + 1}}\sqrt{{\tt 2 L + 1}}$ 
\begin{verbatim} 
                       w9j(L2,L1,L,J2,J1,J,S2,S1,S)     
\end{verbatim} 
\end{center} 
which can be re--written as 
\begin{eqnarray} 
\label{rec-1}
   &   & \hspace*{-4.0cm}  
   \sprm{((L_1,L_2)L,(S_1,S_2)S)J}{((L_1,S_1)J_1,(L_2,S_2)J_2)J} 
   \nonumber \\[0.2cm] 
   & = & 
   \sqrt{[J_1,J_2,L,S]} \, 
   \ninejm{L_1}{S_1}{J_1}{L_2}{S_2}{J_2}{L}{S}{J}
\end{eqnarray} 
by using the symmetries of the Wigner $9-j$ symbols and the fact that 
2J + 2 mJ\_ can be added without any change in the overall phase of the
expression. Using similar lines, we can evaluate the second coefficient
(\ref{general-rec-2}) for the given case of four open ($jj-$coupled) subshells
\spr{((J_{m1},J_{p1})\ J_1,\ (J_{m2},J_{p2})J_2)\ J}{ 
    (((J_{m1},J_{p1})\ J_1,\  J_{m2})J^{\prime}_{12},\ J_{p2})\ J} 
in expression (\ref{exapmle-b})  
\begin{flushleft} 
{\tt 
$ > $ rcc\_2 := Racah\_set(recoupling(`<((Jm1,Jp1)J1,(Jm2,Jp2)J2)J| \\
                \hspace*{7.8cm} (((Jm1,Jp1)J1,Jm2)J12\_p,Jp2)J>`)): \\ 
$ > $ rcc\_2 := Racah\_evaluate(rcc\_2): \\ 
$ > $ Racah\_print(rcc\_2): 
\\[-0.2cm]  
} 
\end{flushleft} 
\vspace{-0.2cm} 
\begin{center} 
\begin{verbatim} 
---> 
               (-Jp2 + 2 mJ_ - J1 + 2 Jm1 + 2 Jp1 - J - Jm2 + 2 J2) 
            (-1) 
\end{verbatim} 
$\sqrt{{\tt 2 J12\_p + 1}}\sqrt{{\tt 2 J2 + 1}}$ 
\begin{verbatim} 
                             triangle(J1,Jm1,Jp1)
                          w6j(J,J1,J2,Jm2,Jp2,J12_p) 
\end{verbatim} 
\end{center} 
which leads to the result 
\begin{eqnarray} 
\label{rec-2}
   &   & \hspace*{-3.0cm}  
   \sprm{((J_{m1},J_{p1})J_1,(J_{m2},J_{p2})J_2)J}{ 
        (((J_{m1},J_{p1})J_1, J_{m2})J^{\prime}_{12},J_{p2})J} 
   \nonumber \\[0.2cm] 
   & = & 
   (-1)^{J_{m2}+J_{p2}+J_1+J} \, \sqrt{[J_2,J^{\prime}_{12}]} \, 
   \sixjm{J_1}{J_{m2}}{J^{\prime}_{12}}{J_{p2}}{J}{J_2} \; .
\end{eqnarray} 

\medskip

When we combine the two expression (\ref{rec-1}) and (\ref{rec-2}), we 
arrive at the expression (\ref{exapmle-explicit}) for the 
transformation matrix of configuration states with two (open) shells in 
$LS-$coupling; their complete expansion in terms of a $jj-$coupled basis 
can be written
\small
\begin{eqnarray} 
\label{exapmle-g} 
   & & \hspace*{-1.5cm} 
   \ketm{(l_1^{\,N_1} \alpha_1 L_1 S_1,\, l_2^{\,N_2} \alpha_2 L_2 S_2) LS\, J} 
   \nonumber \\[0.2cm] 
   & = & 
   \sum_{N_{1_{-}}N_{2_{-}}J_{1}J_{2}J^{\prime}_{12} 
         \tmm{\nu}_1 \tmm{J}_1\tpm{\nu}_1 \tpm{J}_1 
         \tmm{\nu}_2 \tmm{J}_2\tpm{\nu}_2 \tpm{J}_2 } 
   \sqrt{ \left[J_{1},J_{2},L,S\right] } 
   \left\{ \begin{array}{ccc} 
         L_{1} & S_{1} & J_{1} \\ 
         L_{2} & S_{2} & J_{2} \\ 
         L     & S     & J 
   \end{array}  \right\} 
   \nonumber \\[0.2cm] 
   &  \hspace*{1.0cm} \times & 
   (-1)^{\tmm{J}_2+\tpm{J}_2 +J_{1}+J} 
   \sqrt{ \left[J_{2},J^{\prime}_{12},L,S\right] } 
   \left\{ \begin{array}{ccc} 
          J_{1}     & \tmm{J}_2 & J^{\prime}_{12} \\ 
          \tpm{J}_2 & J         & J_{2} 
   \end{array}  \right\} 
   \nonumber \\[0.2cm] 
   &  \hspace*{1.0cm} \times & 
   \sprm{(\tmm{\kappa}_1^{\tmm{N}_1} \tmm{\nu}_1 \tmm{J}_1, \ 
          \tpm{\kappa}_1^{(N_1-\tpm{N}_1)} \tpm{\nu}_1 \tpm{J}_1) J}{ 
          l_1^{N_1} \; \alpha_1 L_1 S_1\ J_1} 
   \nonumber \\[0.2cm] 
   &  \hspace*{1.0cm} \times & 
   \sprm{(\tmm{\kappa}_2^{\tmm{N}_2} \tmm{\nu}_2 \tmm{J}_2, \ 
          \tpm{\kappa}_2^{(N_2-\tpm{N}_2)} \tpm{\nu}_2 \tpm{J}_2) J}{ 
          l_2^{N_2} \; \alpha_2 L_2 S_2\ J_2} 
   \nonumber \\[0.2cm] 
   &  \hspace*{1.0cm} \times & 
   \ketm{(((\tmm{\kappa}_1^{\tmm{N}_1} \tmm{\nu}_1 \tmm{J}_1,         \, 
            \tpm{\kappa}_1^{\tpm{N}_1} \tpm{\nu}_1 \tpm{J}_1) J_1,    \, 
            \tmm{\kappa}_2^{\tmm{N}_2} \tmm{\nu}_2 \tmm{J}_2) J^{\prime}_{12}, \, 
            \tpm{\kappa}_2^{\tpm{N}_2} \tpm{\nu}_2 \tpm{J}_2) J }     \, 
\end{eqnarray} 
\normalsize
or vice versa, if a summation is carried out over all intermediate angular
momenta in $LS-$coupling.

\section{Summary and outlook} 
 
A set of additional commands to the \textsc{Racah} program now facilitates 
the transformation of symmetry--adapted functions with quite different 
complexity from $jj-$ to $LS-$coupling and \textit{vice versa}. For such 
transformations, all partially filled (sub--) shells with $l \le 3$ (i.e.\ 
up to $f-$electrons) are supported and, hence, the program extends the
previously available tabulations and implementations considerably. 
In the study of atomic spectra, for example, the new version of the
\textsc{Racah} program  may help identify atomic and ionic levels as obtained 
from relativistic calculations in $jj-$coupling. Apart from the analysis of the
valence--shell spectra, a reliable classification of the level structure
is crucial, in particular, for the study of inner--shell processes, where 
the creation of additional vacancies often gives rise to a large
(or even huge) number of possible states; in practise, however, only a 
very few levels are typically involved in some process but need first to be
\textit{recognized}, of course. The present extension to the  \textsc{Racah} 
program can help to implement such transformations also directly into 
available atomic code, a project which is currently under work for the 
\textsc{Ratip} package \cite{Ratip}.
 
\medskip 
 
With the implementation of rather abstract data structure such as shell and 
configuration states [cf.\ the auxiliarity procedures \texttt{shell\_{}LS(), 
csf\_{}LS(), ...}], we also provide a powerful notation for more advanced 
tasks. For open--shell atoms and ions, for instance, a long--standing problem 
concerns the computation of the \textit{angular coefficients} for effective 
$n-$particle operators as they occur in many--body perturbation theory. Here,
the given notation for shell and configuration states can help decompose
general matrix elements automatically. Another task concerns the
\textit{optimal classification} of atomic levels to assist the interpretation
of atomic data and to improve the data base on energy levels and transition
probabilities for the large user community of atomic data. For this aim, 
further coupling schemes need to be implemented in the future. 

\vspace*{0.4cm}

\textbf{Acknowledgement:} We like to thank T.\ Inghoff for valuable
suggestions and for help with the installation of the code.

%
%
%
%
%
%
%
%
%
%
%
%
%
%
\section*{Appendix A: Recurrence relations for the subshell transformation
                      coefficients} 
 
The transformation coefficients
\spr{l^{\,N} \; \alpha LS\ J}{(\tmm{\kappa}^{\tmm{N}} \tmm{\nu} \tmm{J}, \ 
                               \tpm{\kappa}^{\tpm{N}} \tpm{\nu} \tpm{J}) J}  
between the (sub--)shell states from different coupling schemes are the 
\textit{building blocks} for all transformations. For a given single--shell
configuration of $N$ equivalent electrons $(l^N)$, these matrices can be 
expressed recursively in terms of the transformation matrices for $N-1$ 
equivalent electrons, i.e.\ in terms of the transformation of
the corresponding parent states. The recurrence relations for the subshell
transformation coefficients therefore include the coefficients of fractional 
parentage 
$ \left(l^{\,N}\; \alpha LS  \left\|  
        l^{\,N-1}\; (\alpha' L' S')\ l \right. \right) \, $ 
and a proper recoupling of the angular momenta
\begin{small} 
\begin{eqnarray} 
\label{q:theory-c} 
   & & \hspace*{-1.5cm} 
   \sprm{l^{N} \; \alpha LS\ J}{(\tmm{\kappa}^{\tmm{N}} \tmm{\nu} \tmm{J}, \ 
                                 \tpm{\kappa}^{\tpm{N}} \tpm{\nu} \tpm{J}) J} 
   \nonumber \\[0.2cm] 
   & = &  
   \sqrt{\left[L,S\right]/N} \, 
   \sum_{\alpha' L' S'}  
   \left(l^{N}\; \alpha LS  \left\|  
         l^{N-1}\; (\alpha' L' S')\ l \right. \right) \, 
   \sum_{J'} \, [J'] 
   \nonumber \\[0.2cm] 
   &   & \hspace*{0.3cm} \times  
   \left[ \, 
      \sqrt{\tmm{N} [j_-,\tmm{J}]} 
      \ninejm{L'}{l}{L}{S'}{s}{S}{J'}{j_-}{J} \, 
      \sum_{\tmm{\nu}' \tmm{J}'} \,  
      (-1)^{j_- +\tmm{J}-\tpm{J}+J'} 
      \sixjm{\tpm{J}}{\tmm{J}'}{J'}{j_-}{J}{\tmm{J}} 
   \right.  
   \nonumber \\[0.2cm] 
   &   & \hspace*{0.7cm} \times 
   \left( \left. j_-^{(\tmm{N}-1)} \; 
      (\tmm{\nu}' \tmm{J}')\ j_-  \right\| 
      j_-^{\tmm{N}_-} \; \tmm{\nu} \tmm{J} \right) \, 
   \sprm{ l^{N-1} \; \alpha' L' S'\ J'} 
      {(\tmm{\kappa}^{(\tmm{N}-1)} \tmm{\nu}' \tmm{J}', \  \tpm{\kappa}^{\tpm{N}}  
      \tpm{\nu} \tpm{J}\ ) J' } 
   \nonumber \\[0.2cm] 
   &   & \hspace*{0.3cm} +  
   \left. \, 
      \sqrt{\tpm{N} [j_-,\tpm{J}]} 
      \ninejm{L'}{l}{L}{S'}{s}{S}{J'}{j_+}{J} \, 
      \sum_{\tpm{\nu}' \tpm{J}'} \, (-1)^{j_{+}+\tmm{J}+\tpm{J}'+J} \, 
      \sixjm{J'}{\tmm{J}}{J'}{J}{j_+}{\tpm{J}}   \right.  
   \nonumber \\[0.2cm] 
   &   & \hspace*{0.7cm} \times  
   \left. 
   \left( \left. j_{+}^{(\tpm{N} -1)} \; 
      (\tpm{\nu}' \tpm{J}') j_{+} \right\| 
      j_{+}^{\tpm{N}} \; \tpm{\nu} \tpm{J} \right) 
   \sprm{l^{(N-1)}\; \alpha' L' S'\ J'} 
      {(\tmm{\kappa}^{\tmm{N}} \tmm{\nu}\tmm{J}, \ \tpm{\kappa}^{(\tpm{N}-1)} 
      \tpm{\nu}' \tpm{J}') \ J' } \right] \; .
\end{eqnarray} 
\end{small} 
They can be applied to the transformation of any subshell state by starting 
from 
\begin{small} 
\begin{eqnarray} 
\label{q:theory-d} 
   \sprm{l^{\,2} \; \alpha LS\ J}{(\tmm{\kappa} \tmm{\nu} \tmm{J}, \ 
                                   \tpm{\kappa} \tpm{\nu} \tpm{J})\ J} 
   & = &  
   \frac{1}{\sqrt{2}} \, \left( 1+(-1)^{L+S}\right)\,  
   \sqrt{[j_{-},j_{+},L,S]} \, 
   \ninejm{l}{l}{L}{s}{s}{S}{j_{-}}{j_{+}}{J} 
\end{eqnarray} 
\end{small} 
and
\begin{small} 
\begin{eqnarray} 
\label{q:theory-f} 
   \sprm{l^{\,2} \; \alpha LS\ J}{\tpmm{\kappa}^{\,2} \tpmm{\nu}\ J} 
   & = &  
   \frac{1}{4} \, \left(1+(-1)^{L+S}\right) \,  
                  \left(1+(-1)^{ J }\right) \, [j_{\pm}] \, \sqrt{L,S]} \, 
   \ninejm{l}{l}{L}{s}{s}{S}{j_{\pm}}{j_{\pm}}{J} \; .
\end{eqnarray} 
\end{small} 

In the present work, the recurrence relations (\ref{q:theory-c}) have been 
utilized to generate the transformation matrices for all partially filled 
shells with $l \le 3$, i.e.\ up to $f-$electrons, and for occupation 
numbers $N \,=\, 1,\ 2,\ ...,\ 2l+1$. In this process, we made use 
of the coefficients of fractional parentage 
$\left(l^{N}\; \alpha LS \left\| l^{N-1}\, (\alpha' L' S')\ l \right. \right)$  
in $LSJ$--  and 
$\left. \left(j^{N-1}\; (\nu' J')j \right\| j^{N}\; \nu J\right)$ 
in $jj-$coupling, which were implemented earlier into the \textsc{Racah}  
package \cite{Gaigalas:01}. To keep the (current) transformation of the
symmetry--adapted functions feasible in time, these
coefficients are stored internally in the program [cf.\ section 4]. For all
occupation numbers larger than $2l+1$, i.e.\  $N \,=\, 2l+2,\  ...,\ 4l+1$,
we make use of the symmetry relation \cite{Dyall_Grant}  
\begin{eqnarray} 
\label{RelLSjj} 
   & & \hspace*{-1.5cm} 
   \sprm{l^{\,N}\; \alpha\nu LS\,J} 
   {({\tmm{\kappa}}^{\tmm{N}} \tmm{\nu}\tmm{J}, \ 
   {\tpm{\kappa}}^{\tpm{N}} \tpm{\nu}\tpm{J})\ J }  
   \nonumber \\[0.2cm]  
   & = & 
   (-1)^{(\nu-\minus{\nu}-\plus{\nu})/2} 
   \sprm{l^{\,4l+2-N}\; \alpha\nu LS\,J } 
   {{\tmm{\kappa}}^{(2j_{_{-}}+1-{\tmm{N}})} \tmm{\nu}\tmm{J}, \ 
   {\tpm{\kappa}}^{(2j_{_{+}}+1-{\tpm{N}})} \tpm{\nu}\tpm{J})\, J } \; . 
\end{eqnarray} 
which is easily derived from the following two properties of the coefficients 
of fractional parentage [cf.\ Eq.\ (15) in Ref.\ \cite{method5} and 
Eq.\ (9) in Ref.\ \cite{method7}]
\begin{eqnarray} 
\label{RelCFPLS} 
   & & \hspace*{-1.5cm} 
   \left( \left. l^{\,4l+1-N}\; (\alpha^{\prime} \nu^{\prime}
                                      L^{\prime}S^{\prime})l 
   \right\| l^{\,4l+2-N}\; \alpha \nu LS\right) 
   \nonumber \\[0.2cm] 
   & = &  
   (-1)^{S+S^{\prime}+L+L^{\prime}-l-\frac{1}{2}+
   \frac{1}{2}(\nu + \nu^{\prime}-1)} 
   \nonumber \\[0.2cm] 
   &   \hspace*{1.0cm} \times &  
   \left( 
   \frac{(N+1)(2L^{\prime}+1)(2S^{\prime}+1)}{(4l+2-N)(2L+1)(2S+1)} 
   \right)^{\frac{1}{2}} 
   \left(l^{\,N}\; (\alpha^{\prime} \nu^{\prime}
                         L^{\prime}S^{\prime})l \left\| l^{\,N+1}\; 
   \alpha \nu LS \right. \right)
\end{eqnarray} 
and 
\begin{eqnarray} 
\label{RelCFPjj} 
   & & \hspace*{-1.5cm} 
  \left(j^{2j-N}\; (\nu^{\prime}J^{\prime})j \left\| j^{2j+1-N}\; \nu J \right. \right)  
  \nonumber \\[0.2cm] 
  & = & 
  (-1)^{J+J^{\prime}-j+\frac{1}{2}(\nu+\nu^{\prime}-1)} 
  \left( 
  \frac{(N+1)(2J^{\prime}+1)}{(2j+1-N)(2J+1)} 
  \right)^{\frac{1}{2}} 
  \left(j^{N}\; (\nu^{\prime}J^{\prime})j \left\| j^{N+1}\; 
  \nu J \right. \right) \; . 
\end{eqnarray} 
For further details about the properties of the cfp coefficients 
and the subshell transformation matrices (\ref{subshell-coefficients}),  
see Refs.\ \cite{method5,method7} and Ref.\ \cite{Gaigalas/ZR:02}, respectively.

\section*{Appendix B: New commands for the \textsc{Racah} package} 
 
The commands of the present extension to the \textsc{Racah} program can be 
described fairly independent from previous parts. Below, we briefly explain 
those procedures which have been \textit{added} and which are of interest for  
an interactive use of the $LS-jj$ transformation matrices. This provides
a short description of the input and output of the procedures to facilitate also
the understanding of our examples in sections 4 and 5; as previously, we 
follow the style of the former \textit{Maple Handbook} \cite{Redfern:96}. 
A more detailed description of all the presently available commands of the 
\textsc{Racah} package  (at user's level) is distributed with the source code 
in the file \texttt{Racah-commands.ps}. 
 
\medskip 
 
As introduced earlier in the text, the terms of a \textit{subshell state} 
and a \textit{configuration state function}  play a key role in the 
transformation of (coupled) states and the evaluation of matrix elements for  
open shells. They form the \textit{basic entities} in dealing with such  
tasks and are often used to describe the input and output of (many) commands. 
To facilitate the handling of these 'atomic states' 
(i.e.\ the communication with and among the procedures of the \textsc{Racah} 
program), we first introduce a number of auxiliarity procedures for these 
coupled states in $LS-$ and $jj-$coupling. Although several tests are made on  
the particular input of these procedures, they basically return 
\textit{unevaluated} and, thus, serve mainly for keeping necessary information 
together. As seen from their names and list of parameters, these auxiliarity  
procedures are designed so that further \textit{coupling schemes} can be easily 
added later as according requirements arise. Since these procedures frequently 
occur during input and output and no actual manipulation is made, also  
\textit{no} prefix \texttt{Racah\_{}} has been added.

\subsection*{B.1 Auxiliarity procedures} 
 
Presently, procedures for the notation of shell--states and (atomic) CSF are 
provided in $jj-$ and $LS-$coupling.

\begin{itemize} 
\item \proc{csf\_{}jj(shell\_{}jj)} 
 
  Auxiliarity procedure to represent a configuration state function which 
  is built from a single $jj-$coupled subshell state \ket{\kappa^{\,N} \nu J} 
  or \ket{n\kappa^{\,N} \nu J}. 
 
  \procout   An unevaluated call to \procref{csf\_{}jj(shell\_{}jj)} is 
  returned. 
 
  \procargop  
  (shell\_{}jj,\textit{check}) to check, in addition, that the given 
  quantum numbers in shell\_{}jj() give rise to a valid $jj-$coupled  
  subshell state with $j \, = \, 1/2,\,\ldots,\,7/2$. 
  \map 
  (shell\_{}jj$_1$,shell\_{}jj$_2$,J) to represent a configuration state 
  function (CSF) of two $jj-$coupled subshell states 
  \ket{(\kappa_1^{\,N_1} \nu_1 J_1,\, \kappa_2^{\,N_2} \nu_2 J_2) J} or 
  \ket{(n_1 \kappa_1^{\,N_1} \nu_1 J_1,\, n_2 \kappa_2^{\,N_2} \nu_2 J_2) J}. 
  \newline
  \map 
  (shell\_{}jj$_1$,shell\_{}jj$_2$,J$_{12}$,shell\_{}jj$_3$,J) to represent a  
  configuration state 
  function (CSF) of three $jj-$coupled subshell states 
  \ket{((\kappa_1^{\,N_1} \nu_1 J_1,\, \kappa_2^{\,N_2} \nu_2 J_2) J_{12},\, 
  \kappa_3^{\,N_3} \nu_3 J_3) J } or \newline
  \ket{((n_1 \kappa_1^{\,N_1} \nu_1 J_1,\, n_2 \kappa_2^{\,N_2} 
  \nu_2 J_2) J_{12},\, n_3 \kappa_3^{\,N_3} \nu_3 J_3) J}. \newline
  \map 
  (shell\_{}jj$_1$,shell\_{}jj$_2$,J$_{12}$,shell\_{}jj$_3$,J$_{123}$,shell\_{}jj$_4$,J) 
  to represent a configuration state 
  function (CSF) of four $jj-$coupled subshell states \newline
  \ket{(((\kappa_1^{\,N_1} \nu_1 J_1,\, \kappa_2^{\,N_2} \nu_2 J_2) J_{12},\, 
  \kappa_3^{\,N_3} \nu_3 J_3) J_{123},\, \kappa_4^{\,N_4} \nu_4 J_4) J } or 
  \newline
  \ket{(((n_1 \kappa_1^{\,N_1} \nu_1 J_1,\, n_2 \kappa_2^{\,N_2} \nu_2 J_2) J_{12},\, 
  n_3 \kappa_3^{\,N_3} \nu_3 J_3) J_{123},\, n_4 \kappa_4^{\,N_4} \nu_4 J_4) J}.   
  \newline
  \procadd  All (given) quantum numbers in the parameter list must evaluate 
  to proper integers or half--integers. 
 
  \procsee  \procref{csf\_{}LS()}, \procref{shell\_{}jj()}, 
            \procref{Racah\_{}csf\_{}print()}. 
\item \proc{csf\_{}LS(shell\_{}LS,J)} 
 
  Auxiliarity procedure to represent a configuration state function which 
  is built from a single $LS-$coupled subshell state \ket{l^{\,N} \nu L S J}  
  or \ket{nl^{\,N} \nu L S J}. 
 
  \procout   An unevaluated call to \procref{csf\_{}LS(shell\_{}LS,J)} is 
  returned. 
 
  \procargop  
  (shell\_{}LS,J,\textit{check}) to check, in addition, that the given 
  quantum numbers in shell\_{}LS() and J give rise to a valid $LSJ-$coupled  
  subshell state with $l \, = \, 0,\,\ldots,\,3$. 
  \map 
  (shell\_{}LS$_1$,shell\_{}LS$_2$,L,S,J) to represent a configuration state 
  function (CSF) of two $LS-$coupled subshell states 
  \ket{(l_1^{\,N_1} \nu_1 L_1 S_1,\, l_2^{\,N_2} \nu_2 L_2 S_2) L S J} or 
  \ket{(n_1 l_1^{\,N_1} \nu_1 L_1 S_1,\, n_2 l_2^{\,N_2} \nu_2 L_2 S_2) L S J}. 
   
  \procadd  All (given) quantum numbers in the parameter list must evaluate 
  to proper integers or half--integers. 
 
  \procsee  \procref{csf\_{}jj()}, \procref{shell\_{}LS()}, 
            \procref{Racah\_{}csf\_{}print()}. 
\item \proc{shell\_{}jj(kappa,N,nu,J)} 
 
  Auxiliarity procedure to represent a $jj-$coupled subshell state  
  \ket{\kappa^{\,N} \nu J} for \newline  
  $j \, = \, 1/2,\,\ldots,\,7/2$. 
 
  \procout   An unevaluated call to \procref{shell\_{}jj(kappa,N,nu,J)} is 
  returned. 
 
  \procargop  
  ([n,kappa],N,nu,J) to represent a $jj-$coupled subshell state \newline 
  \ket{n\kappa^{\,N} \nu J}. 
  \map 
  (kappa,N,nu,J,\textit{check}) to check, in addition, that the given 
  quantum numbers give rise to a valid $jj-$coupled subshell state;  
  the program terminates with an proper \textsc{error} message if this  
  is not the case. 
   
  \procadd   
  All quantum numbers (except of $n$) must evaluate to proper integers  
  or half--integers. 
  \map 
  The \textit{relativistic} angular momentum quantum number 
  $\kappa \,=\, \pm \,(j+1/2)$ for $l \,=\, j \pm 1/2$. 
  \map 
  The principal quantum number $n$ is often not required for the  
  transformation of subshell states but enters the notation, if different 
  subshell states are coupled to each other or, in particular, in the  
  evaluation of (most physical) matrix elements. 
  \map 
  All occupation numbers must be in the range $N \, = \,0,\,\ldots,\,(2j+1)$. 
  \map 
  For $N \,\equiv\, 0$, an (unphysical) subshell angular momentum  
  $ j\,=\, -1/2$ is formally allowed in order to facilitate the input for  
  several procedures from Appendix B.2. 
 
  \procsee  \procref{csf\_{}jj()}, \procref{shell\_{}LS()}, 
            \procref{Racah\_{}shell\_{}print()},
            \procref{Racah\_{}tabulate()}.
\item \proc{shell\_{}LS(l,N,nu,L,S)} 
 
  Auxiliarity procedure to represent a $LS-$coupled subshell state  
  \ket{l^{\,N} \nu L S} for \newline $l \, = \, 0,\,\ldots,\,2$. 
 
  \procout   An unevaluated call to \procref{shell\_{}LS(l,N,nu,L,S)} is 
  returned. 
 
  \procargop  
  (l,N,w,nu,L,S) to represent a $LS-$coupled subshell state  \newline 
  \ket{l^{\,N} w \nu L S} for $l \, = \,3$ and the additional quantum number 
  $w \, = \,0,\,\ldots,\,10$. \newline
  \map 
  ([n,l],N,nu,L,S) or ([n,l],N,w,nu,L,S) to represent the $LS-$coupled  
  subshell states \ket{nl^{\,N} \nu L S} or \ket{nl^{\,N} w \nu L S}, 
  respectively. 
  \map 
  (l,N,nu,L,S,\textit{check}) to check, in addition, that the given 
  quantum numbers give rise to a valid $LS-$coupled subshell state  
  with $l \, = \, 0,\,\ldots,\,2$; the program terminates with an proper 
  \textsc{error} message if this is not the case. 
   
  \procadd   
  All quantum numbers (except of $n$) must evaluate to proper integers  
  or half--integers. 
  \map 
  The principal quantum number $n$ is often not required for the  
  transformation of subshell states but enters the notation, if different 
  subshell states are coupled to each other or, in particular, in the  
  evaluation of (most physical) matrix elements. 
  \map 
  All occupation numbers must be in the range $N \, = \,0,\,\ldots,\,2(2l+1)$. 
 
  \procsee  \procref{csf\_{}LS()}, \procref{shell\_{}jj()}, 
            \procref{Racah\_{}shell\_{}print()},
            \procref{Racah\_{}tabulate()}.
 
\end{itemize} 
 
\vspace*{1.5cm} 
\subsection*{B.2 Commands for $LS-jj$ transformations} 
  
\medskip 
 
This Appendix lists the commands for the transformation of coupled (subshell  
and configuration) states where we utilize the (auxiliarity) notation from 
the previous part. This enables us with a very compact but still flexible  
notation for the input and output of the individual procedures; for example,  
a notation like \texttt{...,shell\_{}LS$_a$,shell\_{}LS$_b$,...} means that  
the user may type explicitly  
\texttt{...,shell\_{}LS(l$_a$,N$_a$,nu$_a$,L$_a$,S$_a 
         $),shell\_{}LS(l$_b$,N$_b$,nu$_b$,L$_b$,S$_b$),...} 
in the parameter list or first assign these (unevaluated) calls to 
\procref{shell\_{}LS()} to any variables, say \texttt{wa,\,wb}, and later  
only use these variables at input time: \texttt{...,wa,wb,...}. 
To 'extract' the quantum numbers from these unevaluated calls, the command 
\procref{Racah\_{}tabulate()} is used. 
 
\begin{itemize} 
\item \subproc{Racah\_{}csf\_{}print(csf\_{}jj)} 
 
  Returns a string of type  
  "$\vert$...('$\kappa$$_1$\^{}N$_1$, nu$_1$, J$_1$';  
              '$\kappa$$_2$\^{}N$_2$, nu$_2$, J$_2$')  
  J$_{12}$; ...$>$" to facilitate the printout of $jj-$coupled CSF. 
  The value of $\kappa _i$ is printed in spectroscopic notation such as 
  d\_3/2, f\_7/2, ... \,; if, moreover, the principal quantum number $n$
  is given, a string like 3d\_5/2\^{}2, ... is returned.
 
  \procout   A string is returned. 
 
  \procargop  
  (csf\_{}LS) to return  
  "$\vert$...('l$_1$\^{}N$_1$, nu$_1$, \^{}2S$_1$+1, L$_1$'; 
              l$_2$\^{}N$_2$, nu$_2$, \^{}2S$_2$+1, L$_2$') L$_{12}$, S$_{12}$; ...$>$". 
  The values of l$_i$ and L$_i$ are printed in spectroscopic 
  notation such as s, p, d, ... and S, P, D, ..., respectively. For
  $f-$electrons ($l_i = 3$), the additional quantum number w$_i$ is printed
  in parenthesis such as 
  "$\vert$...(l$_1$\^{}N$_1$, (w$_1$) nu$_1$, \^{}2S$_1$+1, L$_1$; 
              l$_2$\^{}N$_2$, (w$_2$) nu$_2$, \^{}2S$_2$+1, L$_2$) 
                                                  L$_{12}$, S$_{12}$; ...$>$". 
 
  \procsee  \procref{csf\_{}LS(), csf\_{}jj() and Racah\_{}shell\_{}print()}.

\item \proc{Racah\_{}matrix\_{}LS\_{}jj(shell\_{}LS,shell\_{}jj$_- 
                                                 $,shell\_{}jj$_+$,J)} 
 
  Returns the $LS-jj$ transformation matrix 
  \spr{l^{\,N} \nu L S J}{( \tmm{\kappa}^{\,\tmm{N}} \tmm{\nu} \tmm{J},\, 
                            \tpm{\kappa}^{\,\tpm{N}} \tpm{\nu} \tpm{J} ) J} 
  for all subshells with $l \, = \, 0,\,\ldots,\,3$ and the according 
  $ \tmm{\kappa} \,=\, l$ and $ \tpm{\kappa} \,=\, -(l+1)$. 
 
  \procout   A (floating--point) number is returned. 
 
  \procargop  
  (shell\_{}LS,shell\_{}jj$_-$,shell\_{}jj$_+$,J,\textit{algebraic}) to return 
  the same element of the $LS-jj$ transformation matrix but in algebraic form. 
  \newline 
  \map 
  (shell\_{}LS,shell\_{}jj$_-$,shell\_{}jj$_+$,J,\textit{prime}) to return 
  the same element of the $LS-jj$ transformation matrix but in prime--number 
  representation. \newline 
  \map 
  (shell\_{}LS$_1$,shell\_{}LS$_2$,L,S,shell\_{}jj$_{1-}$,shell\_{}jj$_{1+} 
  $,J$_1$,shell\_{}jj$_{2-}$,J$_{12}$,shell\_{}jj$_{2+}$,J) to return \newline
  the  $LS-jj$ transformation matrix \newline 
  {\small 
  \spr{(l_1^{\,N_1} \nu_1 L_1 S_1,\, l_2^{\,N_2} \nu_2 L_2 S_2) L S J}{ 
     ((( \tmm{\kappa}_1^{\,\tmm{N}_1} \tmm{\nu}_1 \tmm{J}_1,         \, 
         \tpm{\kappa}_1^{\,\tpm{N}_1} \tpm{\nu}_1 \tpm{J}_1  ) J_1,  \, 
         \tmm{\kappa}_2^{\,\tmm{N}_2} \tmm{\nu}_2 \tmm{J}_2  ) J_{12},\, 
         \tpm{\kappa}_2^{\,\tpm{N}_2} \tpm{\nu}_2 \tpm{J}_2  ) J }  } 
    
  \procadd  
  The subshell angular momenta and occupation numbers are not independent 
  of each other; they must fulfill the relation  
  $l \,=\, j_- +1/2 \,=\, j_+ -1/2$ and $N \,=\, \tmm{N} + \tpm{N}$; 
  the program terminates with an proper \textsc{error} message if this is not 
  the case. 
  \map 
  If the principal quantum number $n$ is given, it must be the same for all 
  (sub--) shells. 
  \map 
  For two and more \textit{coupled} subshell states, these relations and 
  condition must hold for each group of subshell states on the lhs and rhs  
  of the transformation matrix. 
  \map 
  For $l \,=\, 0$ follows $\tmm{N} \,=\, 0$ and $\tmm{\kappa} \,=\, 0$; 
  such (unphysical) subshell states are formally allowed in the  
  \textsc{Racah} program but can also be omitted from the list of parameters 
  above. 
  \map 
  The subshell shell\_{}jj$_-$ with $\tmm{\kappa}$ or 
  shell\_{}jj$_+$ with $\tpm{\kappa}$ can be omitted from the list of parameters 
  above if $\tmm{N}=0$ or $\tpm{N}=0$. 
  \map 
  For details about the prime--number representation, see  
  \procref{Racah\_{}calculate\_{}prime()}. 
 
  \procsee  \procref{Racah\_{}set\_{}coupling()}. 
\item \proc{Racah\_{}set\_{}coupling(\textit{LS})} 
 
  Defines the use of $LS-$coupling for the \ket{l^{\,N} \nu L S}  
  subshell states. 
 
  \procout   A \textsc{null} expression is returned. 
 
  \procargop  
  (SL) to defines the use of $SL-$coupling for the \ket{l^{\,N} \nu S L}  
  subshell states. 
   
  \procadd  
  The information about the current coupling scheme for the $LS$ subshell  
  states is kept in the global variable  
  \procref{Racah\_{}save\_{}coupling\_{}LS}; its default value 
  is \procref{Racah\_{}save\_{}coupling\_{}LS=LS}. 
  \map 
  If the coupling scheme of the subshell states is to be changes, this  
  procedure must be called \textit{before} any transformation is made. 
 
  \procsee  \procref{Racah\_{}set\_{}coupling\_{}scheme()}. 
\item \subproc{Racah\_{}shell\_{}print(shell\_{}jj)} 
 
  Returns a string "$\kappa$, \^{}N, nu, J" to facilitate the printout  
  of $jj-$coupled  subshell states. 
  The value of $\kappa _i$ is printed in spectroscopic notation such as 
  d\_3/2, f\_7/2, ... \,; if, moreover, the principal quantum number $n$
  is given, a string like 3d\_5/2\^{}2, ... is returned.
 
  \procout   A string is returned. 
 
  \procargop  
  (shell\_{}jj,\textit{state}) to return "$\vert$$\kappa$, \^{}N, nu, J$>$". 
  \map 
  (shell\_{}LS) to return "l\^{}N, nu, \^{}2S+1, L" or "l\^{}N, w, nu, \^{}2S+1, L". 
  The values of l and L are printed in spectroscopic 
  notation such as s, p, d, ... and S, P, D, ..., respectively. 
  \map 
  (shell\_{}LS,\textit{state}) to return "$\vert$l\^{}N, nu, \^{}2S+1, L$>$" or  
  "$\vert$l\^{}N, w, nu, \^{}2S+1, L$>$". 
   
  \procadd  
  These strings facilitate the line--mode printout of (coupled) subshell states 
  and CSF. 
 
  \procsee  \procref{shell\_{}LS(), shell\_{}jj() and Racah\_{}csf\_{}print()}. 
\item \proc{Racah\_{}tabulate(shell\_{}jj)} 
 
  Return a table with all defined quantum numbers of a $jj-$coupled subshell 
  state. 
   
  \procout   A table \texttt{T} with entries  
  \texttt{T[n], T[kappa], T[N], T[nu],} and \texttt{T[J]} is returned. 
 
  \procargop  
  (shell\_{}LS) to return a table \texttt{T} with all defined quantum numbers  
  of a $LS-$coupled shell state; it has the entries  
  \texttt{T[n], T[l], T[N], T[w], T[nu], T[L],} and \texttt{T[S]}. 
     
  \procadd 
  If some quantum numbers such as the orbital quantum number $l$ is not 
  defined, \textsc{fail} is returned  for the corresponding entry. 
  
  \procsee  \procref{shell\_{}jj()}, \procref{shell\_{}LS()}. 

\item \proc{Racah\_{}transform\_{}asf("jj--$>$LS",csf\_{}jj$_1$,a$_1 
                                                $,csf\_{}jj$_2$,a$_2$,...)} 
 
  Expands an atomic state function, which is represented in a $jj-$coupled  
  CSF basis  
  $$ \ketm{ \Psi_\alpha } \;=\;  
     \sum_{k} \, \ketm{ {\rm CSF}^{\,(jj)}_k} \, a^{\,(jj)}_k (\alpha) \;, $$ 
  into a basis of $LS-$coupled CSF, i.e.\ 
  $$ \ketm{ \Psi_\alpha } \;=\;  
     \sum_{i} \, \ketm{ {\rm CSF}^{\,(LS)}_i} \, c^{\,(LS)}_i  \,.$$ 
 
  \procout   A list [ [csf\_{}LS$_1$,c$_1$], [csf\_{}LS$_2$,c$_2$], ...] 
  is returned where csf\_{}LS$_{\,i}$ describes a $LS-$coupled CSF and 
  c$_i$ the corresponding mixing coefficient in the expansion. 
 
  \procargop  
  ("jj--$>$LS",csf\_{}jj$_1$,a$_1$,csf\_{}jj$_2$,a$_2$,...,\textit{algebraic})  
  to return the mixing coefficients in algebraic form. 
  \map 
  ("jj--$>$LS",csf\_{}jj$_1$,a$_1$,csf\_{}jj$_2$,a$_2$,...,\textit{print})  
  to print the expansion in line mode. 
  One line is printed per term $c_i \,*\, \ketm{ {\rm CSF}^{\,(LS)}_i} $, and 
  a \textsc{null} expression is returned in this case. 
  \map 
  ("LS--$>$jj",csf\_{}LS$_1$,a$_1$,csf\_{}LS$_2$,a$_2$,...) to expand an  
  atomic state function, which is represented in a $LS-$coupled  
  CSF basis  
  $$ \ketm{ \Phi_\alpha } \;=\;  
     \sum_{k} \, \ketm{ {\rm CSF}^{\,(LS)}_k} \, a^{\,(LS)}_k (\alpha) \;, $$ 
  into a basis of $jj-$coupled CSF, i.e.\ 
  $$ \ketm{ \Phi_\alpha } \;=\;  
     \sum_{i} \, \ketm{ {\rm CSF}^{\,(jj)}_i} \, c^{\,(jj)}_i \,.$$   
     
  \procadd  
  The subshell states of all $jj-$coupled CSF must be provided in  
  \textit{standard order}, i.e.\ if both $jj-$subshells with  
  $j \,=\, l \pm 1/2$ occur in a CSF, they must always couple like 
  $$ (\tmm{\kappa}^{\,\tmm{N}} \tmm{\nu} \tmm{J},  
      \tpm{\kappa}^{\,\tpm{N}} \tpm{\nu} \tpm{J} ) J $$ 
  if they represent the first two subshells, and  
  $$ ((...,\tmm{\kappa}^{\,\tmm{N}} \tmm{\nu} \tmm{J} ) J,  \, 
           \tpm{\kappa}^{\,\tpm{N}} \tpm{\nu} \tpm{J} ) J'... )$$ 
  otherwise. 
  \map 
  For any expansion into a $jj-$coupled CSF basis, the subshell states 
  also appear in \textit{standard order} in the output. 
  \map 
  If the principal quantum number(s) $n$ are given, they are transfered
  properly to the output but must be the same for each group of subshell 
  states in the expansion of the CSF.
  \map 
  For the use of a $SL-$coupled CSF basis, see 
  \procref{Racah\_{}set\_{}coupling()}. 
  \map 
  The subshell $\tmm{\kappa}$ or $\tpm{\kappa}$ can be omitted from the 
  arguments of csf\_{}jj if $\tmm{N}=0$ or $\tpm{N}=0$. 
 
  \procsee  \procref{csf\_{}jj(), csf\_{}LS(), shell\_{}jj(), shell\_{}LS()}.

\item \proc{Racah\_{}transform\_{}csf("jj--$>$LS",csf\_{}jj)} 
 
  Expands a $jj-$coupled CSF into a basis of $LS-$coupled CSF 
  $$ \ketm{ {\rm CSF}^{\,(jj)}} \;=\;  
     \sum_{i} \, \ketm{ {\rm CSF}^{\,(LS)}_i} \, c_i \,.$$ 
 
  \procout   A list [ [csf\_{}LS$_1$,c$_1$], [csf\_{}LS$_2$,c$_2$], ...] 
  is returned where csf\_{}LS$_{\,i}$ describes a $LS-$coupled CSF and 
  c$_i$ the corresponding mixing coefficient in the expansion. 
 
  \procargop  
  ("jj--$>$LS",csf\_{}jj,\textit{algebraic}) to return the mixing coefficients  
  in algebraic form. 
  \map 
  ("jj--$>$LS",csf\_{}jj,\textit{print}) to print the expansion in line mode. 
  One line is printed per term $c_i \,*\, \ketm{ {\rm CSF}^{\,(LS)}_i} $, and 
  a \textsc{null} expression is returned in this case. 
  \map 
  ("LS--$>$jj",csf\_{}LS) to expand a $LS-$coupled CSF into a basis of  
  $jj-$coupled CSF 
  $ \ketm{ {\rm CSF}^{\,(LS)}} \;=\;  
     \sum_{i} \, \ketm{ {\rm CSF}^{\,(jj)}_i} \, c_i \,$. 
     
  \procadd  
  The subshell states of all $jj-$coupled CSF must be provided in  
  \textit{standard order}, i.e.\ if both $jj-$subshells with  
  $j \,=\, l \pm 1/2$ occur in a CSF, they must always couple like 
  $$ (\tmm{\kappa}^{\,\tmm{N}} \tmm{\nu} \tmm{J},  
      \tpm{\kappa}^{\,\tpm{N}} \tpm{\nu} \tpm{J} ) J $$ 
  if they represent the first two subshells, and  
  $$ ((...,\tmm{\kappa}^{\,\tmm{N}} \tmm{\nu} \tmm{J} ) J,  \, 
           \tpm{\kappa}^{\,\tpm{N}} \tpm{\nu} \tpm{J} ) J'... )$$ 
  otherwise where $ \tmm{\kappa} \,=\, l$ and $ \tpm{\kappa} \,=\, -(l+1)$. 
  \map 
  For any expansion into a $jj-$coupled CSF basis, the subshell states 
  also appear in \textit{standard order} in the output. \newline 
  \map 
  If the principal quantum number(s) $n$ are given, they are transfered
  properly to the output but must be the same for each group of subshell 
  states in the expansion of the CSF.
  \map 
  For the use of a $SL-$coupled CSF basis, see 
  \procref{Racah\_{}set\_{}coupling()}. 
  \map 
  The subshell $\tmm{\kappa}$ or $\tpm{\kappa}$ can be omitted from 
  the argument of the csf\_{}jj if $\tmm{N}=0$ or $\tpm{N}=0$. 
 
  \procsee  \procref{csf\_{}jj(), csf\_{}LS(), shell\_{}jj(), shell\_{}LS()}.

\item \proc{Racah\_{}transform\_{}csf\_{}jj\_{}LS(csf\_{}jj)} 
 
  Expands a single $jj-$coupled CSF into a basis of $LS-$coupled CSF 
  $$ \ketm{ {\rm CSF}^{\,(jj)}} \;=\;  
     \sum_{i} \, \ketm{ {\rm CSF}^{\,(LS)}_i} \, c_i \,.$$ 
 
  \procout   A list [ [csf\_{}LS$_1$,c$_1$], [csf\_{}LS$_2$,c$_2$], ...] 
  is returned where csf\_{}LS$_i$ describes a $LS-$coupled CSF and 
  c$_i$ the corresponding mixing coefficient in the expansion. 
 
  \procargop  
  (csf\_{}jj,\textit{algebraic}) to return the mixing coefficients in 
  algebraic form. 
   
  \procsee  \procref{csf\_{}jj()}, \procref{csf\_{}LS()}, 
            \procref{shell\_{}jj()}, \procref{shell\_{}LS()}, 
            \procref{Racah\_{}set\_{}coupling()}, \newline
            \procref{Racah\_{}transform\_{}csf()}.
\item \proc{Racah\_{}transform\_{}csf\_{}LS\_{}jj(csf\_{}LS)} 
 
  Expands a single $LS-$coupled CSF into a basis of $jj-$coupled CSF 
  $$ \ketm{ {\rm CSF}^{\,(LS)}} \;=\;  
     \sum_{i} \, \ketm{ {\rm CSF}^{\,(jj)}_i} \, c_i \,.$$ 
 
  \procout   A list [ [csf\_{}jj$_1$,c$_1$], [csf\_{}jj$_2$,c$_2$], ...] 
  is returned where csf\_{}jj$_i$ describes a $jj-$coupled CSF and 
  c$_i$ the corresponding mixing coefficient in the expansion. 
 
  \procargop  
  (csf\_{}LS,\textit{algebraic}) to return the mixing coefficients in 
  algebraic form. 
   
  \procsee  \procref{csf\_{}jj()}, \procref{csf\_{}LS()}, 
            \procref{shell\_{}jj()}, \procref{shell\_{}LS()}, 
            \procref{Racah\_{}set\_{}coupling()}, \newline
            \procref{Racah\_{}transform\_{}csf()}.

\end{itemize} 
 
%
%
%
%
%

\newpage 

%
%
%
%
\section*{TEST RUN OUTPUT} 
 
\begin{small} 
\begin{verbatim} 
 > Racah_LS_jj_calculate_table(); 
"LS-jj transformation matrices for f subshells with occupation N=7"
"J=", 1/2
"f^7, (w=0) nu=5, ^6F"
"f_5/2^1 nu=1, 5/2", "f_7/2^6 nu=2, 2", [1, 2, 1, 0, -3]
"f_5/2^2 nu=2, 2", "f_7/2^5 nu=3, 3/2", [-1, 4, 1, -1, -4]
"f_5/2^2 nu=2, 2", "f_7/2^5 nu=3, 5/2", [-1, 0, 0, -1, -3, 1]
"f_5/2^2 nu=2, 4", "f_7/2^5 nu=1, 7/2", [1, 3, 0, 1, -3]
"f_5/2^2 nu=2, 4", "f_7/2^5 nu=3, 9/2", [-1, 0, 0, 0, -4, 1, 1]
"f_5/2^3 nu=1, 5/2", "f_7/2^4 nu=2, 2", [0]
"f_5/2^3 nu=1, 5/2", "f_7/2^4 nu=4, 2", [1, 0, 1, 0, -3, 1]
"f_5/2^3 nu=3, 3/2", "f_7/2^4 nu=2, 2", [1, 1, 1, 0, -4]
"f_5/2^3 nu=3, 3/2", "f_7/2^4 nu=4, 2", [0]
"f_5/2^3 nu=3, 9/2", "f_7/2^4 nu=2, 4", [1, 2, 0, 2, -4, 1]
"f_5/2^3 nu=3, 9/2", "f_7/2^4 nu=4, 4", [0]
"f_5/2^3 nu=3, 9/2", "f_7/2^4 nu=4, 5", [0]
"f_5/2^4 nu=2, 2", "f_7/2^3 nu=3, 3/2", [-1, 4, 1, -1, -4]
"f_5/2^4 nu=2, 2", "f_7/2^3 nu=3, 5/2", [-1, 0, 0, -1, -3, 1]
"f_5/2^4 nu=2, 4", "f_7/2^3 nu=1, 7/2", [-1, 3, 0, 1, -3]
"f_5/2^4 nu=2, 4", "f_7/2^3 nu=3, 9/2", [-1, 0, 0, 0, -4, 1, 1]
"f_5/2^5 nu=1, 5/2", "f_7/2^2 nu=2, 2", [-1, 2, 1, 0, -3]
.
.
\end{verbatim} 
\end{small} 
 
\end{document}